\documentclass[12pt]{article}
\usepackage[utf8]{inputenc}
\usepackage{epsfig}
\usepackage{amsmath,amsthm,amsfonts, mathtools,dsfont}
\usepackage{soul,color}
\usepackage{multirow}
\usepackage{multicol}
\usepackage{bm, bbm}
\usepackage{epsfig}
\usepackage{natbib}
\usepackage{latexsym, graphicx, alltt}
\usepackage{graphicx}
\usepackage{setspace}
\usepackage{rotating}
\usepackage{threeparttable}
\usepackage{lscape}
\usepackage{geometry}
\usepackage{booktabs}
\usepackage{comment}
\usepackage{longtable}
\usepackage[colorlinks=true,linkcolor=blue,urlcolor=blue,citecolor=blue,hyperfootnotes=false]{hyperref}
\usepackage[table]{xcolor}
\usepackage{tikz}
\usepackage{pgfplots}
\pgfplotsset{compat=1.18}

\renewcommand{\baselinestretch}{1.2}
\newcommand{\al}{\alpha}
\newcommand{\be}{\beta}
\newcommand{\ga}{\gamma}

\renewcommand{\th}{\theta}

\newcommand{\convd}{\stackrel{d}\rightarrow}

\usepackage{nicematrix,tikz}
\usetikzlibrary{matrix, fit}

\pgfkeys{tikz/mymatrix/.style={matrix of math nodes,inner sep=0pt,row sep=0em,column sep=0em,nodes={inner sep=6pt}}}
\usetikzlibrary{decorations.pathreplacing}

\newtheorem{example}{Example}

\def\baselinestretch{1.2}
\usepackage{bm}
\usepackage{epsfig}
\begin{document}

\title{Grouped fixed effects regularization for binary choice models}
\author{Claudia Pigini\footnote{Marche Polytechnic University
    (Italy). E-mail: \url{c.pigini@univpm.it}} \and Alessandro
  Pionati\footnote{Marche Polytechnic University (Italy).Corresponding
    Author. Address: Department of Economics and Social Sciences, P.le
    Martelli 8, 60121 Ancona (Italy) E-mail:
    \url{a.pionati@univpm.it}} \and Francesco Valentini
  \footnote{University of Pisa (Italy). E-mail:
    \url{francesco.valentini@unipi.it}}}

\date{}
\maketitle

\def\baselinestretch{1.2}  % spacing --  no details in royal A

\thispagestyle{empty}%to hide the page number of the titlepage
 %abstract w/o referencing for submission 
 %max 200 words
 
\begin{abstract}
We study the application of the grouped fixed effects approach to binary choice models for panel data in presence of severe complete separation. Through data loss, complete separation may lead to  biased estimates of Average Partial Effects and imprecise inference. Moreover, forecasts are not available for units without variability in the response configuration. The grouped fixed effects approach \emph{discretizes} unobserved heterogeneity via k-means clustering, thus reducing the number of fixed effects to estimate. This regularization reduces complete separation, since it relies on within-cluster rather than within-subject response transitions. Drawing from asymptotic theory for the APEs, we propose choosing a number of groups such that clustering delivers a good approximation of the latent trait while keeping the incidental parameters problem under control. The simulation results show that the proposed approach delivers unbiased estimates and reliable inference for the APEs. Two empirical applications illustrate the sensitivity of the results to the choice of the number of groups and how nontrivial forecasts for a much larger number of units can be obtained.
\end{abstract}

\small
  \vskip 3mm
  \noindent {\bf Keywords:} {\sc Average Partial Effects, Dynamic models, Grouped fixed effects,\\ Rare events, Regularization } %alph order

\setcounter{page}{0}  % reset page counter 

\clearpage
\section{Introduction}\label{sec:intro}
Fixed-effects (FE) binary choice models are prominently used in applied
econometrics and popular examples arise from a wide range of 
applications.\footnote{Noteworthy examples come from labor
market participation \citep{heckman1980does} with a focus on fertility choices for
female married workers \citep{hyslop1999state}, self-reported health
status \citep{contoyannis2004dynamics}, transitions in income dynamics
\citep{cappellari2004modelling}, household finance 
\citep{alessie2004ownership}, and drivers of unionization choices
\citep{wooldridge2005simple}. Recent applications can be found in
studies on firms' behavior in accessing credit
\citep{pigini2016state}, migrants’ remitting choices
\citep{bettin2018dynamic},
link formation models \citep{dzemski2019},
energy poverty \citep{drescher2021determinants},persistence of innovation in
firms \citep{arroyabe2022estimation}.} Estimation of model parameters in this context, where one or more sets of FE are included, is usually carried out by Maximum Likelihood (ML). 

With binary panel data, it may happen that the dependent variable does not exhibit within-subject variation when outcomes describe highly persistent phenomena or extremely rare events, such as employment status and the occurrence of financial crises. Practitioners estimating FE binary choice models based on these data are often unable to recover finite estimates of all the individual intercepts, an issue known in the literature as the complete separation (CS) problem \citep{albert1984existence}: for a given subject, if their outcome configuration does not vary over time, the log-likelihood
will be monotone in their intercept, leading to a non-finite ML estimate of their FE. Although the true population intercept is finite, the observed time series might not be long enough to observe a time-varying configuration.\footnote{To give the dimension of the CS problem in typical settings, \cite{kunz2021predicting}
 describes an application on health care utilization where 29\% to 45\% of subjects do not exhibit outcome variation over time.
  In the study on labor market participation in \cite{dj2015} and \cite{fernandez2009fixed}, revisited here in Section \ref{sec:applic_labor},  60\% of the observations are in CS.} 

Statistical software typically remove subjects in CS, which, in absence of cross-sectional dependence, has no direct effects on the regression parameter estimates. The sub-sampling, however, impacts other quantities of interest in three main respects: i) the Average Partial Effects (APEs) are overestimated, as the subjects dropped from the dataset are likely to have small individual population partial effects, due to their large index functions; ii) as APEs converge at the rate $1/\sqrt{N}$, with $N$ being the number of subjects, the reduced sample size leads to an imprecise large sample approximation of the APE sampling distribution, resulting in poor finite-sample coverage; iii) forecasts for discarded units are trivial, as their predicted probability would always be zero or one, in and out of sample.

This paper motivates the application of the Group Fixed Effects (GFE) approach, put forward by \cite{blm2022}, in settings where the use of FE leads to pervasive CS. The GFE approach is based on
a two-step procedure: in the first step individual, possibly continuous, 
unobserved heterogeneity (UH) is discretized by {\em
k-means} clustering based on the model covariates; in the second step, group-membership indicators enter the main specification as cluster-specific intercepts. The intrinsic regularization introduced by GFE, which limits the number of FE to be estimated, reduces the instances of CS. This happens because the existence of finite estimates for the group-specific intercepts  relies on the within-cluster, as opposed to within-subject, variability in outcome configuration. Therefore, subjects without outcome variation are retained if they end up in a cluster together with individuals who exhibit time variability in their response configuration.

We show that the GFE regularization effectively overcomes 
the finite-sample issues entailed by the CS-related sample size reduction: i) APEs computed using GFE estimates account for the systematically smaller marginal effects of subjects otherwise dropped, providing a more precise quantification of the population APEs; 
ii) the larger sample size actually used yields more accurate coverage for the APEs; iii) the GFE approach allows one to make non-trivial predictions for units without variation in the response variable, as long as these are clustered in groups where outcome variation at cluster-level is observed.

Ways of dealing with CS are the subject of the stream of literature that relies on shrinkage to obtain finite ML estimates. These approaches are inspired by the modified score correction introduced by \cite{firth1993bias} for the logit model, applied to handle CS in cross-section data by \cite{Heinze2002} and \cite{Heinze2006}, then generalized by \cite{Kosmidis2009} to nonlinear models of the exponential family. Modified versions of this approach have later been used to shrink FE estimates in binary choice models by \cite{kunz2021predicting} and \cite{pigini2021penalized}, who focus on forecasts, and \cite{Cook2018}, who suggest FE shrinkage to reliably quantify population APEs by means of a plug-in estimator. Despite the conjecture put forward by \cite{Cook2018}, a thorough study of finite-sample properties nor complete asymptotic theory for APEs with a Firth-type shrinkage is available. For instance, it is well known in the literature that plug-in APE estimators still suffer from the typical incidental parameters problem, which might not be negligible when the individual time series is short, thus requiring a bias correction \citep{dj2015}. 

Further to providing evidence of better coverage of the GFE plug-in APE estimator, we show that the cluster regularization employed by the proposed approach can be used to limit the effects of the incidental parameters problem on the APE estimator in finite samples. Relying on the asymptotic properties of the proposed estimator, we provide the practitioner with a guideline to choose a number of groups that simultaneously makes the incidental parameters bias and the approximation error entailed by discretization both negligible in finite samples. Therefore, no further bias reduction is required. Finally, it is worth to stress that the GFE approach can directly be applied to dynamic binary choice models, differently from the shrinkage-type estimators that would require a modification of the score correction term.

%_______SIMULATION__________________________
The simulation study analyzes the finite-sample properties of the GFE plug-in estimator of the APE, for both static and dynamic logit models, in presence of moderate to severe degrees of CS. The results show that the GFE approach mitigates the APEs overestimation, which would otherwise result from dropping subjects in CS, as witnessed by the performance of the infeasible estimator. The performance of the proposed approach is also compared to the APEs plug-in estimators obtained using ML and to the analytical and jackknife bias-corrected APE estimators \citep{hahn2011bias,dj2015}. By discarding significantly fewer observations, the GFE APE estimator exhibits minimal bias and better empirical coverage. Moreover, choosing the number of groups approximating individual UH according to the proposed rule makes the incidental parameters bias negligible in finite samples, signaling that further bias reduction can be avoided.\footnote{We also explore the performance of the plug-in estimator based on the Firth-type score correction \citep{firth1993bias}. While this approach does not lead to loss of observations, the shrinkage of the FE estimates does not seem to be effective as a bias reduction device.}

\begin{comment}
provide imprecise estimates for APE, with misleading inference based on poor coverage. In particular, in the static case we show how CS leads to the overestimation of APE.\footnote{Despite being present also in the dynamic setting, the overestimation of APE does not emerge in simulation due to the opposite effect of Nickell’s bias, which leads to a compensation, see Section \ref{sec:sim} for details.}
\end{comment}

%________APPLICATIONS
We  present the results of two real-data applications.  The first  revisits the empirical application on the participation of young working women in the labor market proposed, among others, by \cite{dj2015} and \cite{fernandez2009fixed}. In this setting, CS involves around 60\% of the original sample due to the strong intertemporal correlation of employment status, a phenomenon often observed in labor market studies. We show that, as in the simulation study, the GFE approach retains a larger portion of the dataset and leads to a quantification of APEs that coherently lies between the pooled and the ML-based bias-corrected estimators. 
The second application presents a forecast exercise based on rare events. We use the panel data on financial crises issued by \cite{laeven2018systemic}, where the dependent variable is equal to one if a country in a particular year witnessed financial turmoil. We show that the GFE approach manages to offer non-zero predicted probabilities for a higher number of countries with respect to ML alternatives and has a good forecasting performance. 

\begin{comment}
 Overall, the simulation study and the empirical application suggest the use of GFE approach when the degree of CS is particularly high.
\end{comment}

The rest of the paper is organized as follows: Section \ref{sec:ec_meth} outlines the effects of CS in ML estimation of FE binary choice models and motivates the use of the GFE approach; Section \ref{sec:sim} presents the simulation study; Section \ref{sec:applic} illustrates the two empirical applications. Finally, Section \ref{sec:conclusions} concludes.

\section{Econometric methods}\label{sec:ec_meth}
%___________ ML ___________________________________________
\subsection{Background on fixed-effects binary choice models} \label{sec:background}

For $i=1,\ldots,N$ and $t=1,\ldots,T$, we study the model
\begin{equation}  \label{ass:model_baseline}
y_{it} = \mathbbm{1}(x_{it}' \beta_0 + \alpha_{i0}  + u_{it} > 0),
\end{equation}
where  $ \mathbbm{1}(\cdot)$ is the indicator function, $x_{it}$  denotes a set of $J$ individual-specific covariates associated with a conformable vector of unknown parameters $\beta_0$ and may include  $y_{i,t-1}$; $\alpha_{i0}$ parameterizes the UH as time-invariant individual effects, while $u_{it}$ is the i.i.d error, whose distribution is either standard logistic or normal.
%Moreover, we assume that $N,T \to \infty$ and $N/T \to \rho$ \citep{Li2003}. 

The structural and nuisance parameters in model \eqref{ass:model_baseline} can be jointly estimated using ML, leading to $(\hat{\beta}', \hat{\alpha}_1, \ldots,\hat{\alpha}_N)'$. As is well known, the ML estimator suffers from the so-called incidental parameters problem (IPP), which is due
to the estimation noise introduced by the nuisance parameters entering the profile likelihood for the structural ones \citep{NS1948}. The IPP leads to an asymptotic bias in the limiting distribution, even if both $N$ and $T$ $\to \infty$, but in a fixed proportion to each other.\footnote{This framework is referred to as rectangular array asymptotics \citep{Li2003}, where $N,T \to \infty$ with $N/T \to \rho$, $0 <\rho < \infty$.} Bias reduction techniques for the ML estimator are available, in the form of both analytical \citep{fernandez2009fixed, hahn2011bias} and jackknife \citep{dj2015} corrections.

%______________ APE_______________________
The objects of interest in binary choice models are usually the APEs. Let us define the population APE as
\begin{equation}\label{APE0}
  \mu_0 = \mathbb{E}[\mu_{it}(\beta_0,\alpha_{i0})],
\end{equation}
where $\mu_{it}(\beta_0,\alpha_{i0}) = F'(x_{it}'\beta_0 +
\alpha_{i0}) \beta_{0}$ and $F'(\cdot)$ is the first derivative of the probit/logit link function. The ML plug-in estimator of $\mu_0$ is readily available as
 \begin{equation}\label{APE_ML}
  \hat{\mu}=\frac{1}{NT}\sum_{i}\sum_{t}\mu_{it}(\hat{\beta},\hat{\alpha}_{i}),
  \end{equation}
and its asymptotic expansion is such that $\hat{\mu} = \mu_0 + O_p(1/T)$, where the $O_p()$ term represents the bias arising from IPP \citep{hahn2011bias}. Unlike the ML estimator of $\beta_0$, any plug-in APE estimator does not converge at the rate $1/\sqrt{NT}$, but more slowly, as stated by Theorem 5.1 by \cite{dj2015}. Define 
$\mu_i = T^{-1}\sum_t \mu_{it}(\beta_0,\alpha_{i0})$ and $\sigma^2_\mu = \underset{N \to \infty}{\mathrm{lim}} N^{-1} \sum_{i=1}^N (\mu_i - \mu_0)^2$. Then they show that 
as $N,T \to$ $\infty$ with $N/T \to \rho$, $0 <\rho < \infty$, we have:
\begin{equation*}
\sqrt{N}(\hat{\mu} - \mu_0) + O_p\left(\frac{1}{\sqrt{T}}\right) \convd N(0,\sigma^2_\mu).
\end{equation*}
The above expression clarifies that the plug-in ML APE estimator converges at the rate $1/\sqrt{N}$ and the IPP bias, now captured by the term $O_p(1/\sqrt{T})$, is now asymptotically negligible, as it vanishes as $T \to \infty$. However, 
this bias may still be present in finite samples, especially when the observed time series is short or the IPP is particularly severe (e.g., Nickell's bias in dynamic models). Therefore, the use of analytical or jackknife bias corrections is advised  for APEs \citep{fernandez2009fixed, dj2015}.

%_________________________
%___ CS and its consequences _______________________-

Our main concern in this context is the 
 FE estimate $\hat{\alpha}_i$ in finite samples.
\begin{comment}
Define $y_i=(y_{i1},\ldots, y_{iT})$ as the vector for the response variable configuration of individual $i$. 
\end{comment}
Whenever $\sum_{t=1}^T y_{it}=0$ or $T$, meaning that there is no variability in the dependent variable, the ML estimate of $\alpha_{i0}$ does not exist finite, which is an instance of CS.
\begin {example}
As an example, consider a static FE logit model without covariates: it is easy to see that the individual likelihood  $\ell_i= \al_i\sum_{t}^{T}y_{it} - T\log\left[1+\exp(\al_i)\right]$ is maximized at $   \hat{\al}_i=\log(\frac{p_i^*}{1 - p_i^*})$, where $ p_i^*=\sum_{t}^{T} y_{it}/T$. Therefore, the ML estimate of the individual intercept is not finite when $p_i^*$ is either $0$ or $1$.
\end {example}
Statistical software usually removes subjects in CS from the dataset. Although this reduction  has no impact on the estimates of structural parameters $\beta_0$ in absence of cross-sectional dependence, the quantities computed using the predicted probabilities exhibit a bias that depends on the intensity of the CS problem. 
\begin{comment}
 
Consider the predicted probabilities:
  \[
p_{it} =F\left(x_{it}'\beta_0 + \alpha_{i0}\right)
\]
where $F(\cdot)$ is the probit/logit link function. 
\end{comment}
Non-finite estimates of $\alpha_{i0}$ lead to an estimated probability $F(x_{it}'\hat{\beta} + \hat{\alpha}_{i})$ exactly equal to zero or one. 
%, and the removal of problematic units causes the overestimation of the plug-in ML estimator of the APE in equation \eqref{APE_ML}. 
Consider expression \eqref{APE_ML} in presence of CS:
\[
\widehat{\mu}^\ast = \frac{1}{N^\ast T} \sum_{i \not\in D}\sum_{t}\mu_{it}(\hat{\beta},
\hat{\alpha_i}),
\]
where $D=\{ i: \sum_{t}^Ty_{it} = 0 \,\ \text{or} \sum_{t}^Ty_{it} = T \}$ is the subset of individuals for whom no transitions are observed in the outcome variable and $N^\ast < N$ the number of individuals who are not in CS. Data reduction causes the overestimation of $\mu_0$, because discarded units tend to have a large index functions in absolute value and, in turn, small individual population partial effects, usually close to zero. After the removal of problematic units, the distribution of the estimated PEs (in absolute value) becomes left-truncated as the smaller values are excluded. Because $\widehat{\mu}^\ast$ is computed using only the PEs of individuals who do not belong to $D$, the APE, conditional on this restricted sample, is systematically greater than $\mu_0$. The resulting quantification of the effects of interest is then imprecise.

\begin{example}
  Consider a static logit model including a single binary explanatory
  variable $x_{it}$ and $T=2$. The PE for a generic individual $i$ is
\[
  PE_i(\alpha_{i0}, \beta_0)= F(\alpha_{i0}+ \beta_0)-F(\alpha_{i0}).
\]
% Define
% $s_i = 1\{ y_{i1}\neq y_{i2} \}$, so that $s_i=1$ identifies
% individuals who are \emph{not dropped} from the estimation
% sample.
Consider the case in which
$x_i = \{x_{i1},x_{i2}\} = \{0,1\}$. Conditional on $x_i$, the probability of not observing a
change in outcomes is
\begin{gather*}
  P(i \in D \mid\alpha_{i0}) = P(y_{i1} = 0, y_{i2} = 0 \mid\alpha_{i0}) + P(y_{i1} = 1, y_{i2} = 1 \mid\alpha_{i0}) = \\
  [1-F(\alpha_{i0})][1-F(\alpha_{i0}+\beta_0)]
  + F(\alpha_{i0})F(\alpha_{i0}+\beta_0). 
\end{gather*}
\begin{figure}[h!b]
\caption{
Probability of CS and PE
}
\label{fig:example_cs}
\centering
\begin{tikzpicture}
  \begin{axis}[
    width=11cm, height=6cm,
    domain=-6:5.5,
    samples=600,
    xlabel={$\alpha_{i0}$},
    ylabel={},
    ymin=0, ymax=1,
    axis lines=left,
    axis line style={thick},
    tick style={thick},
    grid=both,
    major grid style={gray!20},
    clip=false, % permette alla legenda di uscire dai limiti
    legend style={
      at={(0.5,1.05)},
      anchor=south,
      legend columns=1,
      draw=none,
      fill=none,
      font=\small,
      inner ysep=10pt 
    }
  ]

  % funzione logistica
  \pgfmathdeclarefunction{F}{1}{\pgfmathparse{1/(1+exp(-#1))}}

  % P(s=1 | alpha)
  \pgfmathdeclarefunction{psel}{1}{%
    \pgfmathparse{F(#1)*(1-F(#1+1)) + (1-F(#1))*F(#1+1)}%
  }

  % Delta(alpha)
  \pgfmathdeclarefunction{Delta}{1}{%
    \pgfmathparse{F(#1+1)-F(#1)}%
  }

  % (1 - P(s=1))
  \addplot [very thick, red!70!black, dashed] ({x},{1 - psel(x)});
  \addlegendentry{$P(i \in D \mid\alpha_{i0}, \; \beta_0 = 1)$ }

  % Delta(alpha)
  \addplot [very thick, blue!70!black] ({x},{Delta(x)});
  \addlegendentry{$PE_i(\alpha_{i0}) = F(\alpha_{i0}+1)-F(\alpha_{i0})$}

  % annotazioni
  \node[anchor=south west, font=\small, color=blue!70!black] at (axis cs:-0.7,0.25)
        {};
  \node[anchor=south east, font=\small, color=red!70!black] at (axis cs:-5,0.95)
        { };
  \node[anchor=south east, font=\small, color=red!70!black] at (axis cs:3,0.95)
        {};

  \end{axis}
\end{tikzpicture}
\end{figure}
In Figure \ref{fig:example_cs}, we illustrate $PE_i(\alpha_{i0})$ and
the probability of being dropped
$P(i \in D \mid \alpha_{i0})$ for $\beta_0=1$.
%
% 
% The analytical expressions simplify to
% \[
% \begin{aligned}
% APE_i(\alpha_{i0}) &= F(\alpha_{i0}+1) - F(\alpha_{i0}),
% P(s_i=0 \mid \alpha_{i0}) &= 1 - \frac{e^{\alpha_{i0}}(1+e)}{(1+e^{\alpha_{i0}})(1+e^{\alpha_{i0}+1})}.
% \end{aligned}
% \]
As we 
can see, individuals who are more likely to be dropped due to
complete separation correspond precisely to those with extreme fixed
effects and negligible partial effects.
\end{example}

Heavy data separation has an additional effect on the estimation of the APEs. Since $\hat{\mu}$ converges to the real APE at the rate $N^{-1/2}$, finite sample performance of the estimator crucially relies on the availability of a large number of individuals. However, unless $N$ is large, when events are extremely rare or outcomes are very persistent, $N^\ast \ll N$ might be too small for the asymptotic
approximation to deliver good coverage. In this vein, $\hat{\mu}^\ast$ can also lead to misleading inference. 

Finally, ML estimation in presence of CS leads to trivial forecasts for discarded subjects, potentially giving rise to misclassification instances, regardless of the threshold $\tau \in [0,1]$ used to build the test-set confusion matrix. In certain contexts, this is not without consequences: one example is that of rare events (low-probability, high impact). In fact, the predicted probability for a subject will be non-zero only if another event has been experienced by the same unit in the past, thus preventing a meaningful forecast of a first-ever occurrence. 

\begin{comment}
Consider, for instance, a scenario in which the rare event is observing $y_{it}=1$, meaning that the ML estimates $\hat{\alpha_i} = - \infty$ for all units for which $\sum_{t}^Ty_{it} = 0$.
In formulae:
\[
\hat{p}_{it}=\mathbb{I}\left[F(x_{it}'\hat{\beta} - \infty ) > \tau\right] = 0 \,\,
\forall \,\, t \,\, \mathrm{and} \,\, \tau,
   \]   
where $\tau \in [0,1]$ is a cut-off point. Predictions for units in CS are tied to their past history and lead to biased outcomes whenever the dependent variable changes for the first time in the year of forecast. At the same time, discarding problematic units is often not an option, as analysts  require  forecast for every unit in the dataset, given also that in presence of rare events the percentage of dropped units is substantial. 
\end{comment}

%___________GFE ______________________________________________
\subsection{Grouped fixed-effects estimation with complete separation}

In this Section, we illustrate how the use of the GFE approach \citep{blm2022} mitigates the issues arising with CS by limiting the number of subjects dropped due to the lack of outcome variation.

The GFE approach is based on the idea that individual UH $\al_{i0}$ can be approximated by a smaller set of group-specific parameters. 
Grouped structures of heterogeneity, which are assumed to be discrete in the population, are becoming increasingly popular in the FE literature \citep{hahnmoon2010,bm2015,lumsdaine2023estimation,mugnier2025simple}. In contrast, \cite{blm2022}'s approach is in the same spirit of contributions that employ clustered structures to \emph{approximate} general forms - both continuous and discrete - of UH
\citep{beyhum2024inference, freemanweidner2023} and it is, to the best of our knowledge, the only viable for nonlinear models.
%____ PROCEDURE________________________________
The GFE estimation procedure consists of two steps:
\begin{enumerate}
\item {\bf Classification step} The individual heterogeneity $\alpha_{i0}$ is {\em discretized} by {\em kmeans} clustering, which uses the vector of the $J$ individual averages $\bar{x}_{i} = T^{-1}\sum_{t} x_{it}$. The algorithm partitions individuals into $K$ groups, with $K \ll N$, such that
\begin{equation}\label{eq:kmeans}
(\bar{x}_{\hat{k}=1}, \ldots, \bar{x}_{\hat{k}=K},\hat{k}_1, \ldots, \hat{k}_N) = \mathrm{argmin} \sum_{i=1}^N || \bar{x}_i - \bar{x}_{k_i} ||^2,
\end{equation}
where $\bar{x}_{k}$ is the mean of $\bar{x}_i$ in group $k$.

\item {\bf Estimation step} Consists of the ML estimation of the model 
\[
y_{it} = \mathbbm{1}(x_{it}' \beta_0 + \alpha_{\hat{k}_i}  + u_{it} > 0),
\]
where $\alpha_{\hat{k}_i} = \alpha_k \mathbbm{1}(i \in k)$, $k = 1,
\ldots,K$. These are the cluster-specific FE, related to the group-membership dummies, and are estimated jointly with the structural parameters yielding
$(\tilde{\beta}', \tilde{\alpha}_1, \ldots, \tilde{\alpha}_K)'$.
\end{enumerate}
Regularity conditions for the validity of {\em kmeans} clustering  and details on the asymptotic properties of the GFE estimator are given in \cite{blm2022, suppblm2022}.

%_____________ MOMENTS W/o Y _________________________
It is worth highlighting that the moments used for the {\em kmeans} clustering have to be informative about the UH: for $T \to \infty$, \cite{blm2022} clarify that they have to be functions of the UH such that one can separate two individuals with a different level of UH by comparing their vector of moments.\footnote{In particular, Assumption 2 in \cite{blm2022} requires moments to be injective. Denote the individual UH as $\xi_{i0}$, of unspecified form, such that $\alpha_{i0} = \alpha(\xi_{i0})$, where $\alpha(\cdot)$ is a Lipschitz-continuous function. Then there exist moments $h_i = (1/T) \textstyle{\sum_t} h(y_{it},x_{it}')$ and a Lipschitz-continuous function $\phi(\cdot)$ such that $\mathrm{plim}_{T \to \infty} h_i = \phi(\xi_{i0})$; moreover, there exist a Lipschitz-continuous function $\psi(\cdot)$ such that $\xi_{i0} = \psi^{-1}(\phi(\xi_{i0}))$.}  In principle, not only the moments of the regressors, but also $\bar{y}_i = T^{-1} \sum_{t=1}^T y_{it}$ can be used. However, when outcomes are highly persistent or very rare, $\bar{y}_i$ may exhibit small to no variability between subjects when the time series is short, which hampers its ability to inform us about different types of UH. For this reason, in our subsequent simulation study and empirical applications, we avoid using the individual averages of the dependent variable in clustering.    

% ______ CS Reduction _____________________________________
The smaller number of FE to estimate decreases the likelihood of dealing with a set of individuals
clustered in a group where no variability in the outcome variable is
observed. In practice, if individuals with no transitions are clustered together with non-problematic
ones, the ML estimate of the related group-specific intercept exists finite. As a consequence, the number of observations discarded due to CS is lower.
\medskip

\noindent
%_________EXAMPLE___________________
\textbf   {Example 1 (continued)}\emph{
Consider again the example about the static panel logit model without covariates. Applying the GFE procedure leads to three possible solutions for $\alpha_{\hat{k}}$:
\[
   \tilde{\alpha}_{k} = \log\left(\frac{p_{\hat{k}}}{1 - p_{\hat{k}}}\right) =
  \begin{cases}
    - \infty & \text{if} \quad p_{\hat{k}}= 0, \\
    \text{finite} & \text{if} \quad p_{\hat{k}} \in (0,1),\\
    \infty & \text{if} \quad p_{\hat{k}}= 1,
  \end{cases} 
\]
where  $p_{\hat{k}} = \frac{\sum_{i \in k} \sum_{t} y_{it}}{T\sum_{i}\mathbbm{1}(i \in k)}$ is the average of the outcomes in 
group $k$. In Table \ref{illex}, we consider a panel composed by
$N=2$ individuals and $T=2$ time periods. The second individual does
not show any state transition, therefore it is not possible to have a finite estimate for $\hat{\al}_{2}$. If, instead, the two individuals are clustered in the same group ($k=1$), within-cluster variability allows us to obtain a finite ML estimate of their shared intercept $\tilde{\alpha}_{1}$.}\footnote{In the example $p_{1}=3/4$ and $\tilde{\al}_{1}=\log{3}\approx 1.1$.} 
\begin{table}
\caption{Illustrative example on CS problem}
\label{illex}
\begin{center}
\begin{tabular}{lcccccc}
  \hline
  \\
 & id & time & $y_{it}$ &$\hat{\al}_{i}$ & $k$ &$\tilde{\al}_{k}$ \\ [2pt]
  \hline \\
&1&1&1&0&  1   & 1.1 \\[ 3pt]
&1&2&0&&   & \\[ 3pt]
&2&1&1&$\infty$& 1   & 1.1 \\[ 3pt]
&2&2&1& & & \\[ 3pt]
&$\vdots$&$\vdots$&$\vdots$&$\vdots$&$\vdots$&$\vdots$ \\
   \hline 
\end{tabular}
\end{center}
\vspace{-0.4cm}
\centering
\begin{scriptsize}
``id''  refers to individual identifier, ``time'' refers to the time period, \\
``$y_{it}$'' is the observed outcome variable, ``$\hat{\al}_{i}$''  is the ML estimate of the individual FE, \\
``$k$'' is the group membership, ``$\tilde{\al}_{k}$'' is the estimate of the grouped fixed effect.\\
\end{scriptsize}
\end{table}
\medskip

%________ BENEFITS of GFE_____________________-
By limiting the number of FE to be estimated and relying on an increased sample size, the GFE approach introduces a regularization that mitigates the consequences of CS. Unlike popular contributions, such as \cite{phillips2016} and \cite{wang2024panel}, which focus on regularization as a way to credibly identify latent grouped patterns, GFE regularization pertains to the estimation of a  more parsimonious model.

% K --->groups--- rule-----> gamma
The amount of information loss depends on the within-cluster variability implied by the classification step, which makes the choice of the number of groups $K$ crucial in this context. At the same time, in the framework put forward by \cite{blm2022} clustering is an approximation device for an unspecified form of UH, thus the granularity of the discretization is closely tied to the quality of the approximation. \citeauthor{blm2022} propose a rule for choosing $K$ that reflects such trade-off between accuracy and parsimony. Specifically, they set
\[
K = \underset{K \geq 1}{\mathrm{min}} \{K: \widehat{Q}(K) \leq \gamma \widehat{V}_{\bar{x}} \},
\]
where $\widehat{Q}(K)$ here indicates the {\em kmeans} objective function in \eqref{eq:kmeans} and $\widehat{V}_{\bar{x}}$ is an estimate of the variability between  the moment vectors and the individual UH. In practice, they propose choosing the smallest number of groups such that the variability of  $\bar{x}_i$ with respect to the centroids is less than or equal to the variability of $\bar{x}_i$ with respect to the individual UH. This choice is also governed by the specification of the user-specified hyper-parameter $\gamma$, which is bound in $(0,1]$ and such that smaller values require a larger number of groups in order to make the between-centroids variability smaller or equal than $\widehat{V}_{\bar{x}}$. However, no tuning procedure for $\gamma$ is suggested. We here provide further guidance on the choice of $K$ on the basis of the asymptotic expansion for the APEs, that are the main objects of interest in the present context.

\begin{comment}
The number of groups $K$ is not known in practice and must be determined. To address this, \cite{blm2022} proposed a rule\footnote{Technical details on the rule are available in Appendix XXX.} that selects the smallest $K$ such that the objective function of the \emph{k-means} problem (i.e., the variability of $\bar{x}_i$ with respect to the centroids) is less than or equal to the variability of $\bar{x}_i$ with respect to individual UH, multiplied by a user-specified hyperparameter $\gamma \in (0,1]$.
 The rule is conceived to  balance the  trade-off between the number of groups, that is, the goodness of the approximation of the UH, and the severity of the IPP bias, as is clear by the role of $K$ in Equation \ref{APE_GFE}. In empirical settings, the choice of $\gamma$ is not trivial and \cite{blm2022} does not provide a recommendation on its selection. 
\end{comment}

%__________________________ APE GFE
Consider the GFE plug-in APE estimator $\tilde{\mu}$. Then assuming that suitable regularity conditions hold,\footnote{These conditions are contained in Assumptions 1-3 in \cite{blm2022} and Assumption S1 in \cite{suppblm2022}.} the asymptotic expansion of the APE estimator \citep[cf. Corollary S1,][]{suppblm2022} implies that
as $N,K,T\to \infty$, $K/NT \to 0$, with $N/T \to \rho$,
 \begin{equation}\label{eq:APE_GFE}
 \sqrt{N}\left(\tilde{\mu} - \mu_0\right) + O_p\left(\frac{\sqrt{N}}{K^2} \right) + O_p \left( \frac{1}{\sqrt{T}} \right)  + 
 O_p\left(\frac{K}{T\sqrt{N}}\right) \convd N\left(0,\sigma^2_{\mu}\right).
 \end{equation}
The above expression shows that  $\tilde{\mu}$ has three sources of  bias, represented by the $O_p(\cdot)$ terms: the $O_p(\sqrt{N}/K^2 )$ term arises from the approximation error due to the discretization of the UH using the \emph{kmeans} procedure; 
the $O_p( 1/\sqrt{T})$ term originates as a classification-step IPP bias, due to the use of $N$ averages, $\bar{x}_{i}$, for $NT$ observations; finally, the $O_p(K/T\sqrt{N})$ term represents the second step IPP bias due to the estimation of $K$ cluster-specific intercepts using $NT$ observations.

That all bias terms in expression \eqref{eq:APE_GFE} will be asymptotically negligible is guaranteed by $K$ growing at certain rates in relation to $N$ and $T$. In particular, \cite{blm2022} 
suggest that setting $K$ proportional to or greater than $\mathrm{min}(\sqrt{T}, N)$ guarantees that the clustering approximation bias will be of order $O_p(1/T)$ for the GFE estimator, thus $O_p(1/\sqrt{T})$ in the above expansion. In this respect, it is worth noting that setting $K$  in this way generates a constant bias in the asymptotic distribution of the GFE estimator \citep[Corollary 1,][]{blm2022}, giving rise to a negligible term in the distribution of the APE GFE estimator. In addition, we argue that the second-step IPP may also be asymptotically negligible as long as $K$ is chosen to be smaller than $T\sqrt{N}$. This refinement of the rule 
indirectly suggests how to tune $\ga$, which should be chosen within a range of values yielding $\sqrt{T} \ll K < T\sqrt{N}$.

As the GFE approach limits the number of units dropped from the data, the asymptotic distribution offers a better approximation of the sampling one of $\tilde{\mu}$ in finite samples, with respect to the FE approach, thus providing a more accurate coverage. In addition, $\tilde{\mu}$ is less likely to overestimate $\mu_0$, as more  units, including those with small PE, are retained in the dataset. 
Finally, the GFE approach manages to provide nontrivial predictions for units without transition in the outcome variable, as long as they are clustered in groups where outcome variability is observed. This allows practitioners to get finite predictions for every unit.

\begin{comment}
We propose a simple procedure to select $\gamma$  which is rooted in the asymptotic behavior of the GFE plug-in APE estimator and suggest setting the parameter to get a number of groups for which: i)  $K \gg \sqrt{T}$ to have negligible approximation and first step IPP bias and ii) $K < T\sqrt{N}$ so that second step IPP bias is not severe. 
The refinement of the rule leads to a range of values of $\gamma$ in which the biases are balanced and further choice within the interval is left to the analyst.
In practice, we suggest performing the GFE estimation for different values of $\gamma$ and finding the aptest value of $K$ in the interval.
\end{comment}

\section{Simulation study}\label{sec:sim}
\subsection{Static logit model}\label{sec:sim:st}
%------------ DGP -------------------------------
We study the finite sample performance of the GFE approach by estimating a static logit model in presence of CS. We generate data from the model
\begin{equation*}  
y_{it} = \mathbbm{1}(x_{it,1} \beta_1 + x_{it,2} \beta_2 + \alpha_{i}  + u_{it} > 0),
\end{equation*}
where  $\alpha_{i} \sim N (\nu_{\alpha}, 1)$. The two
regressors are generated as $x_{it,j} =\,\  N(0,1) + \al_i$, for $j= 1,2$, and $\beta_1=\beta_2=1$. The error term $u_{it}$ follows an i.i.d standard logistic distribution. 
%-------- Setting -------------------------
We study panels of $N=(100,200)$ individuals observed for
$T=(8,16)$ time occasions. We control for the degree of CS by setting  $\nu_{\alpha} = 1,2$, with a proportion of subjects without individual outcome variation ranging from 40\% to 80\%. For each scenario, we run 1000 Monte Carlo simulations.

 % ------------------GFE ---------------------------
The number of groups $K$ chosen for the GFE approach is implied by a set of equally spaced values of the hyperparameter $\gamma = (0.1, 0.4, 0.7, 1)$. Larger values of $\gamma$ correspond to fewer groups, with $\gamma=1$ yielding the smallest $K$ and therefore the strongest reduction in the number of units incurring in CS. Each value of $\gamma$ implies that $K$ is within $\sqrt{T}$ and $T\sqrt{N}$ in each scenario considered.

%--------------------Comparisons-------------------------------
We compare the plug-in GFE APE estimator for $x_1$ with an infeasible APE estimator that computes average effects only for subjects with outcome variation over time. We also compare the performance of the GFE approach with four alternative APE estimators: the FE plug-in ML estimator, the analytical and iterated jackknife bias corrected APE estimators by \cite{HN2004}, and the APE estimator that plugs in ML Firth-regularized estimates.

% ------------------What we report------------------------------
Tables \ref{tab:static:nu_2}-\ref{tab:static:nu_1} report the mean and median ratios between the estimated and the real population APE, the APE standard deviation ("S.D."), and the empirical size of a two-sided $t$-test\footnote{We use analytical standard errors obtained via Delta Method.} centered in the population APE at significance levels 0.05 and 0.1  ("p .05" and "p .10"). We also report the percentage of observations removed due to CS and the average number of groups (K) implied by the chosen values of $\gamma$.

%---------------- What we can say ----------------------------
First of all, it is worth noting that the infeasible estimator systematically presents a ratio much greater than one, clearly showing that removing observations in CS unavoidably leads to an overestimation of the APE. Coherently, this bias decreases in the percentage of subjects without variability in the response configuration (denoted by the \% of CS for the ML estimator), which gets smaller as $T$ increases and $\nu_\alpha$ is set to $1$. 

The plug-in ML estimator of the APE does not apparently exhibit an upward bias, as it is likely to be offset by the IPP one, which can still shift the sampling distribution when $T$ is small \citep{dj2015}. Nevertheless, for this estimator, coverage issues arise when the percentage of units in CS is elevated. The upward bias in the APE estimator shows up as soon as the IPP bias is reduced
by either an analytical or jackknife correction, thus also affecting coverage accuracy. Finally, the APE estimator that plugs in ML Firth-regularized estimates shows an unsatisfactorily finite sample performance.\footnote{We find that, in the scenarios considered, the FE regularized estimates are systematically smaller than the true individual intercepts, which clearly leads to larger estimated individual partial effects and to an upward bias in the APE estimator.} 

\begin{comment}
No alternative estimator has comparable performance: the FE estimator has an overall limited bias, but this evidence is due to the IPP bias which has a downward sign; Moreover,  the coverage is not precise. 
As discussed in section \ref{sec:background}, the FE and  the debiased estimators tend to overestimate the APE and the quantification of it is rather inaccurate, especially when $T$ is short and the CS is strong. In addition to this, the empirical size is rather distant from nominal levels, making the inference not reliable.
\end{comment}

The regularization entailed by the GFE approach effectively reduces the instances of complete separation for all the values of $\gamma$, and thus  the number of groups considered. Regarding its finite sample performance, overall  the mean and median ratios display smaller biases with respect to the alternative estimators considered, and the larger number of observations retained help to improve the finite-sample coverage. 

The performance of the GFE estimator sensitively varies with the number of groups considered in the classification step. In fact, the bias of the ratio increases with the value of the hyper-parameter: this is a result of the number of groups yielded by $\gamma$ not being large enough to provide an adequate approximation of the underlying UH distribution, even though the average $K$ across simulations complies with the guidelines to choose the number of groups, i.e, $K > \sqrt{T}$. This is expected in our design, as the UH is normally distributed and its support is, for instance,  approximated only by roughly $6$ to $8$ points when $\gamma = 1$. For this reason, it is advisable to choose a value of $\gamma$ which implies $K \gg \sqrt{T}$.

An increase in bias should also be expected for very small values of the hyper-parameter, as a larger number of groups operating the discretization could give rise to an IPP bias in finite samples. However, this issue does not arise in the scenarios considered as, with $\gamma$ as small as $0.1$, the implied number of GFEs to estimate does not seem to be large enough for such bias to show up prominently. 

Finally, it should be noted that the finite-sample coverage of the GFE estimator does not improve with larger sample sizes. This is likely due to the fact that, on average, the number of groups increases only slightly or remains stable when $N$ doubles, so that the approximation bias does not decay, while the confidence interval shrinks instead. 

\begin{comment}
The proposed approach significantly reduces the number of CS instances, the mitigation increasing with the $\gamma$ parameter. For instance, in Table \ref{tab:static:nu_2} the percentage of observations dropped by the GFE estimator is half the ML one for the highest value of $\gamma$.
In turn, the increased number of observations reverberates in the finite sample performance of the plug-in GFE APE, which exhibits low bias. At the same time, the inference is more precise, as is clear from values of the empirical size reported in the Tables, which are close to the nominal size and almost always in the confidence interval.
\end{comment}

%------TABLES for STATIC SIMULATIONS----------------------
%____________ TABLE 1 _____________________
\begin{table}
\caption{Estimated APE of $x_1$, $\nu_{\alpha}=2$}
\label{tab:static:nu_2}
\vspace{-0.2cm}
\begin{center}
  \begin{tabular}{rccccccc}
  \hline\\
 & Mean ratio & Median ratio & S.D. & p .05 & p .10  & CS & K \\ 
  \hline\\ [3pt]
  & \multicolumn{7}{c}{$N=100, T=8$}  \\ %[ 5 pt] 
   \cline{2-8} \\
  %  $N = 100$ &  &  &  &  &  &  &  \\ [ 5 pt]
   Infeasible & 3.426 &  &  &  &  &  &  \\ 
  ML & 0.974 & 0.970 & 0.009 & 0.120 & 0.178 & 79.355 & - \\ 
  BC & 1.300 & 1.287 & 0.010 & 0.411 & 0.484 & 79.355 & - \\ 
  J & 1.159 & 1.160 & 0.032 & 0.639 & 0.688 & 79.355 & - \\ 
  Firth & 1.735 & 1.705 & 0.013 & 0.804 & 0.850 & 0 & - \\ 
  GFE $\gamma$ = 0.1 & 0.999 & 0.988 & 0.008 & 0.079 & 0.127 & 64.034 & 32.523 \\ 
  GFE $\gamma$ = 0.4 & 1.024 & 1.015 & 0.008 & 0.069 & 0.125 & 51.454 & 12.944 \\ 
  GFE $\gamma$ = 0.7 & 1.042 & 1.036 & 0.008 & 0.071 & 0.113 & 44.505 & 8.404 \\ 
  GFE $\gamma$ = 1 & 1.059 & 1.050 & 0.008 & 0.060 & 0.103 & 40.176 & 6.251 \\[ 8 pt]
  & \multicolumn{7}{c}{$N=200, T=8$}  \\ %[ 5 pt] 
   \cline{2-8} \\
Infeasible & 3.400 &  &  &  &  &  &  \\ 
  ML & 0.981 & 0.979 & 0.006 & 0.108 & 0.181 & 79.480 & - \\ 
  BC & 1.314 & 1.307 & 0.007 & 0.592 & 0.669 & 79.480 & - \\ 
  J & 1.314 & 1.294 & 0.020 & 0.632 & 0.697 & 79.480 & - \\ 
  Firth & 1.763 & 1.753 & 0.009 & 0.951 & 0.964 & 0 & - \\ 
  GFE $\gamma$ = 0.1 & 1.014 & 1.020 & 0.006 & 0.059 & 0.113 & 59.513 & 44.957 \\ 
  GFE $\gamma$ = 0.4 & 1.032 & 1.036 & 0.006 & 0.065 & 0.116 & 45.111 & 15.201 \\ 
  GFE $\gamma$ = 0.7 & 1.050 & 1.048 & 0.006 & 0.064 & 0.124 & 38.774 & 9.369 \\ 
  GFE $\gamma$ = 1 & 1.070 & 1.070 & 0.006 & 0.074 & 0.133 & 34.279 & 6.731 \\  [ 0.75 cm]
  & \multicolumn{7}{c}{$N=100, T=16$}  \\ %[ 5 pt] 
   \cline{2-8} \\
  Infeasible & 2.911 &  &  &  &  &  &  \\ 
  ML & 0.992 & 0.990 & 0.006 & 0.064 & 0.128 & 71.809 & - \\ 
  BC & 1.134 & 1.134 & 0.007 & 0.185 & 0.274 & 71.809 & - \\ 
  J & 1.189 & 1.181 & 0.025 & 0.731 & 0.779 & 71.809 & - \\ 
  Firth & 1.515 & 1.494 & 0.008 & 0.811 & 0.866 & 0 & - \\ 
  GFE $\gamma$ = 0.1 & 1.005 & 0.998 & 0.006 & 0.048 & 0.101 & 56.721 & 37.435 \\ 
  GFE $\gamma$ = 0.4 & 1.018 & 1.016 & 0.006 & 0.050 & 0.098 & 44.940 & 15.724 \\ 
  GFE $\gamma$ = 0.7 & 1.030 & 1.030 & 0.006 & 0.049 & 0.098 & 39.016 & 10.233 \\ 
  GFE $\gamma$ = 1 & 1.045 & 1.037 & 0.006 & 0.053 & 0.110 & 34.948 & 7.715 \\  [ 8 pt]
  & \multicolumn{7}{c}{$N=200, T=16$}  \\ %[ 5 pt] 
   \cline{2-8} \\
  Infeasible & 2.879 &  &  &  &  &  &  \\ 
  ML & 0.995 & 0.993 & 0.005 & 0.079 & 0.141 & 71.727 & - \\ 
  BC & 1.136 & 1.130 & 0.005 & 0.288 & 0.390 & 71.727 & - \\ 
  J & 1.238 & 1.226 & 0.017 & 0.722 & 0.764 & 71.727 & - \\ 
  Firth & 1.516 & 1.504 & 0.006 & 0.968 & 0.978 & 0 & - \\ 
  GFE $\gamma$ = 0.1 & 1.009 & 1.007 & 0.005 & 0.059 & 0.116 & 52.573 & 53.389 \\ 
  GFE $\gamma$ = 0.4 & 1.022 & 1.019 & 0.005 & 0.057 & 0.112 & 38.877 & 18.847 \\ 
  GFE $\gamma$ = 0.7 & 1.033 & 1.030 & 0.005 & 0.064 & 0.128 & 32.941 & 11.632 \\ 
  GFE $\gamma$ = 1 & 1.045 & 1.045 & 0.005 & 0.064 & 0.132 & 28.721 & 8.460 \\ 
\hline   
\end{tabular} 
\end{center}
\begin{footnotesize}
Notes: static logit model. Mean and median of the ratio between estimated and
population APEs. S.D.: standard deviation of estimated APE. p.05 and p.10: empirical size of a two-sided $t$-test at the 0.05 and 0.1 significance level.  CS=\% of dropped observations, K= average number of groups in GFE estimators. 1000 Monte Carlo replications.
\end{footnotesize}
\end{table}
%____________TABLE 3 __________________-
\begin{table}
\caption{Estimated APE of $x_1$, $T=8,16$, $\nu_{\alpha}=1$}
\label{tab:static:nu_1}
\vspace{-0.2cm}
\begin{center}
\begin{tabular}{rccccccc}
  \hline\\
  & Mean ratio & Median ratio & S.D. & p .05 & p .10  & CS & K \\ 
  \hline\\ [3pt]
  & \multicolumn{7}{c}{$N=100, T=8$}  \\ %[ 5 pt] 
   \cline{2-8} \\
  Infeasible & 1.642 &  &  &  &  &  &  \\ 
  ML & 0.988 & 0.985 & 0.012 & 0.088 & 0.157 & 49.833 & - \\ 
  BC & 1.053 & 1.049 & 0.012 & 0.126 & 0.191 & 49.833 & - \\ 
  J & 1.111 & 1.106 & 0.044 & 0.610 & 0.675 & 49.833 & - \\ 
  Firth & 1.172 & 1.172 & 0.013 & 0.341 & 0.423 & 0 & - \\ 
  GFE $\gamma$ = 0.1 & 1.001 & 1.002 & 0.011 & 0.052 & 0.094 & 30.694 & 32.523 \\ 
  GFE $\gamma$ = 0.4 & 1.018 & 1.016 & 0.011 & 0.048 & 0.095 & 20.551 & 12.944 \\ 
  GFE $\gamma$ = 0.7 & 1.030 & 1.031 & 0.011 & 0.046 & 0.087 & 16.469 & 8.404 \\ 
  GFE $\gamma$ = 1 & 1.039 & 1.040 & 0.011 & 0.050 & 0.096 & 13.883 & 6.251 \\  [ 8 pt]
  & \multicolumn{7}{c}{$N=200, T=8$}  \\ %[ 5 pt] 
   \cline{2-8} \\
 Infeasible & 1.649 &  &  &  &  &  &  \\ 
  ML & 0.987 & 0.986 & 0.009 & 0.083 & 0.134 & 50.096 & - \\ 
  BC & 1.053 & 1.052 & 0.009 & 0.131 & 0.218 & 50.096 & - \\ 
  J & 1.156 & 1.156 & 0.030 & 0.621 & 0.673 & 50.096 & - \\ 
  Firth & 1.174 & 1.173 & 0.009 & 0.496 & 0.591 & 0 & - \\ 
  GFE $\gamma$ = 0.1 & 1.002 & 1.000 & 0.009 & 0.047 & 0.103 & 26.814 & 44.957 \\ 
  GFE $\gamma$ = 0.4 & 1.015 & 1.014 & 0.008 & 0.052 & 0.104 & 16.359 & 15.201 \\ 
  GFE $\gamma$ = 0.7 & 1.027 & 1.026 & 0.008 & 0.053 & 0.108 & 12.554 & 9.369 \\ 
  GFE $\gamma$ = 1 & 1.036 & 1.035 & 0.008 & 0.060 & 0.112 & 10.328 & 6.731 \\  [ 0.75 cm]
  & \multicolumn{7}{c}{$N=100, T=16$}  \\ %[ 5 pt] 
   \cline{2-8} \\
Infeasible & 1.491 &  &  &  &  &  &  \\ 
  ML & 0.999 & 0.998 & 0.010 & 0.073 & 0.142 & 38.534 & - \\ 
  BC & 1.026 & 1.026 & 0.010 & 0.095 & 0.152 & 38.534 & - \\ 
  J & 1.091 & 1.089 & 0.037 & 0.706 & 0.743 & 38.534 & - \\ 
  Firth & 1.104 & 1.105 & 0.010 & 0.258 & 0.342 & 0 & - \\ 
  GFE $\gamma$ = 0.1 & 1.004 & 1.002 & 0.010 & 0.061 & 0.113 & 24.543 & 37.435 \\ 
  GFE $\gamma$ = 0.4 & 1.012 & 1.010 & 0.010 & 0.065 & 0.109 & 16.614 & 15.724 \\ 
  GFE $\gamma$ = 0.7 & 1.019 & 1.017 & 0.010 & 0.068 & 0.118 & 13.473 & 10.233 \\ 
  GFE $\gamma$ = 1 & 1.027 & 1.028 & 0.010 & 0.071 & 0.121 & 11.399 & 7.715 \\ [ 8 pt] 
  & \multicolumn{7}{c}{$N=200, T=16$}  \\ %[ 5 pt] 
   \cline{2-8} \\
 Infeasible & 1.490 &  &  &  &  &  &  \\ 
  ML & 0.994 & 0.992 & 0.007 & 0.084 & 0.135 & 38.651 & - \\ 
  BC & 1.021 & 1.019 & 0.007 & 0.092 & 0.164 & 38.651 & - \\ 
  J & 1.094 & 1.096 & 0.025 & 0.702 & 0.753 & 38.651 & - \\ 
  Firth & 1.099 & 1.098 & 0.007 & 0.370 & 0.466 & 0 & - \\ 
  GFE $\gamma$ = 0.1 & 1.001 & 1.000 & 0.007 & 0.065 & 0.114 & 21.122 & 53.389 \\ 
  GFE $\gamma$ = 0.4 & 1.009 & 1.008 & 0.007 & 0.060 & 0.131 & 12.938 & 18.847 \\ 
  GFE $\gamma$ = 0.7 & 1.015 & 1.016 & 0.007 & 0.066 & 0.131 & 9.999 & 11.632 \\ 
  GFE $\gamma$ = 1 & 1.022 & 1.023 & 0.007 & 0.080 & 0.137 & 8.539 & 8.460 \\  
\hline 
\end{tabular}
\end{center}
\begin{footnotesize}
Notes: static logit model. Mean and median of the ratio between estimated and
population APEs. S.D.: standard deviation of estimated APE. p.05 and p.10: empirical size of a two-sided $t$-test at the 0.05 and 0.1 significance level.  CS=\% of dropped observations, K= average number of groups in GFE estimators. 1000 Monte Carlo replications.
\end{footnotesize}
\end{table}

\subsection{Dynamic logit model}\label{sec:sim:dyn}

We also study the finite sample performance of the GFE approach by estimating a dynamic logit model in presence of CS.  
%_____________ DGP________________________
For $i =1,\dots, N$ and $t=1,\dots,T$, we generate the outcome variable as
\begin{equation*}
y_{it} = \mathbbm{1}(\be y_{i,t-1} +  x_{it,1} \theta_1 + x_{it,2} \theta_2 + \alpha_{i}  + u_{it} > 0),
\end{equation*}
where $\th_1=\th_2=1$ and $\be=0.5$.
The two regressors and the time-invariant FE are generated as $ x_{it,j} =\,\ N(0,1) + \al_i$ for $j=1,2$, with $\al_i \sim N(\nu_{\alpha},1)$, respectively. 
%_____________ SETTING _______________________

We study panels of $N=(100, 200)$ individuals observed for $T=(8,16)$ time occasions. We control for the degree of CS with two values of $\nu_{\alpha}=(0,-1)$, which results in a percentage of 24\% to 50\% of units without outcome variation. We run 1000 Monte Carlo simulations for each scenario.
%____________ What is in the tables__________________
We report simulation statistics for the APE estimator of $y_{t-1}$ in Tables \ref{tab:dyn:1}-\ref{tab:dyn:2}. The estimators analyzed and the values of $\gamma$ are the same selected for static design, with two exceptions: (i) we do not include the APE estimator that plugs in ML Firth-regularized estimates, since its employment in dynamic settings lacks a theoretical background, and (ii) we use bias-correction methods suited for dynamic models, namely the analytical one of \cite{fernandez2009fixed} and the half-panel jackknife estimator \citep{dj2015}.

%____________ What we can say_________________________
As in the static case, the infeasible estimator systematically overestimates the population APE with the bias  decreasing in $T$ and in the  values considered for $\nu_{\alpha}$.
 The FE plug-in ML estimator and both the analytical and jackknife bias corrected APE estimators exhibit poor performance. When the dimension $T$ is short, the IPP is severe and, in turn, the overestimation of the APE is offset by a strong downward bias. Bias corrections manage to improve the mean and median ratios as $T$ increases, although the coverage remains overall inaccurate.
 
 The GFE approach preserves its regularizing properties in the dynamic setting  and manages to reduce the instances of CS for all values of $\ga$. However, the finite sample properties of the GFE plug-in estimator suggest that regularization does not fully offset the stronger IPP bias that arises in dynamic settings, causing the APE to be systematically underestimated when $\gamma = 0.1$. Accordingly, this bias decreases with larger $T$. 
However, the bias already decreases and the empirical coverage attains its nominal values with intermediate values of $\gamma$, such as $0.4$ and $0.7$, especially when $T=16$.

Appendix \ref{app:sim} contains additional simulation evidence related to a data generating process that violates the assumption of the stationarity of regressors \citep[cf. Assumption 3(i),]{blm2022}.
Table \ref{tab:trending} shows, however, that the results on the finite sample properties of the proposed estimator are robust to the inclusion of a trending regressor. 
 
%______________| TABLES |____________________________

%_____________ TAB 1 _________________-
\begin{table}
\caption{Estimated APE of $y_{t-1}$,  $\nu_{\alpha}=-1$}
\label{tab:dyn:1}
\begin{center}
\begin{tabular}{rccccccc}
  \hline\\
  & Mean ratio & Median ratio & S.D. & p .05 & p .10  & CS & K \\ 
  \hline\\ [3pt]
& \multicolumn{7}{c}{$N=100, T=8$}  \\ %[ 5 pt] 
   \cline{2-8} \\
   Infeasible & 1.615 &  &  &  &  &  &  \\ 
  ML & -0.383 & -0.410 & 0.022 & 0.757 & 0.836 & 50.6 & -  \\ 
  BC & 0.853 & 0.823 & 0.028 & 0.157 & 0.223 & 50.6 & - \\ 
  J & 0.567 & 0.537 & 0.030 & 0.254 & 0.367 & 50.6 & - \\ 
  GFE $\gamma$ = 0.1 & 0.676 & 0.652 & 0.025 & 0.104 & 0.171 & 30.8 & 32.408 \\ 
  GFE $\gamma$ = 0.4 & 1.040 & 1.027 & 0.026 & 0.060 & 0.118 & 20.6 & 12.872 \\ 
  GFE $\gamma$ = 0.7 & 1.148 & 1.143 & 0.026 & 0.068 & 0.128 & 16 & 8.352 \\ 
  GFE $\gamma$ = 1 & 1.235 & 1.228 & 0.027 & 0.076 & 0.128 & 13.6 & 6.262 \\  [ 8 pt]
& \multicolumn{7}{c}{$N=200, T=8$}  \\ %[ 5 pt] 
   \cline{2-8} \\
Infeasible & 1.621 &  &  &  &  &  &  \\ 
ML & -0.396 & -0.392 & 0.015 & 0.957 & 0.977 & 51 & - \\ 
  BC & 0.833 & 0.843 & 0.020 & 0.172 & 0.250 & 51 & - \\ 
  J & 0.538 & 0.532 & 0.020 & 0.329 & 0.430 & 51 & - \\ 
  GFE $\gamma$ = 0.1 & 0.800 & 0.797 & 0.018 & 0.092 & 0.151 & 26.9 & 44.841 \\ 
  GFE $\gamma$ = 0.4 & 1.088 & 1.085 & 0.019 & 0.070 & 0.115 & 16.6 & 15.184 \\ 
  GFE $\gamma$ = 0.7 & 1.178 & 1.175 & 0.019 & 0.083 & 0.131 & 12.6 & 9.341 \\ 
  GFE $\gamma$ = 1 & 1.238 & 1.229 & 0.019 & 0.087 & 0.148 & 10.4 & 6.739 \\  [0.75cm]
  & \multicolumn{7}{c}{$N=100, T=16$}  \\ %[ 5 pt] 
   \cline{2-8} \\
   Infeasible & 1.479 &  &  &  &  &  &  \\ 
   ML & 0.199 & 0.192 & 0.016 & 0.535 & 0.662 & 39.2 & -\\ 
  BC & 0.907 & 0.896 & 0.019 & 0.093 & 0.147 & 39.2 & - \\ 
  J & 0.807 & 0.786 & 0.020 & 0.160 & 0.241 & 39.2 & - \\ 
  GFE $\gamma$ = 0.1 & 0.746 & 0.744 & 0.018 & 0.108 & 0.180 & 24.8 & 37.497 \\ 
  GFE $\gamma$ = 0.4 & 0.966 & 0.949 & 0.019 & 0.063 & 0.119 & 16.5 & 15.770 \\ 
  GFE $\gamma$ = 0.7 & 1.040 & 1.026 & 0.019 & 0.059 & 0.108 & 13.3 & 10.285 \\ 
  GFE $\gamma$ = 1 & 1.102 & 1.094 & 0.019 & 0.062 & 0.098 & 11.3 & 7.762 \\  [8pt]
& \multicolumn{7}{c}{$N=200, T=16$}  \\ %[ 5 pt] 
   \cline{2-8} \\
  Infeasible & 1.480 &  &  &  &  &  &  \\ 
  ML & 0.225 & 0.224 & 0.011 & 0.771 & 0.851 & 39.4 & - \\ 
  BC & 0.939 & 0.944 & 0.013 & 0.084 & 0.144 & 39.4 & - \\ 
  J & 0.836 & 0.833 & 0.014 & 0.171 & 0.281 & 39.4 & - \\ 
  GFE $\gamma$ = 0.1 & 0.851 & 0.848 & 0.013 & 0.088 & 0.136 & 21.5 & 53.641 \\ 
  GFE $\gamma$ = 0.4 & 1.032 & 1.035 & 0.013 & 0.052 & 0.101 & 13.2 & 18.949 \\ 
  GFE $\gamma$ = 0.7 & 1.096 & 1.092 & 0.013 & 0.058 & 0.121 & 10 & 11.695 \\ 
  GFE $\gamma$ = 1 & 1.134 & 1.134 & 0.013 & 0.059 & 0.125 & 8.5 & 8.554 \\ 
   \hline 
\end{tabular}
\end{center}
\begin{footnotesize}
Notes: dynamic logit model. Mean and median of the ratio between estimated and
population APEs. S.D.: standard deviation of estimated APE. p.05 and p.10: empirical size of a two-sided $t$-test at the 0.05 and 0.1 significance level.  CS=\% of dropped observations, K= average number of groups in GFE estimators. 1000 Monte Carlo replications.
\end{footnotesize} 
\end{table}
%______________ TAB 2
\begin{table}
\caption{Estimated APE of $y_{t-1}$,$\nu_{\alpha}= 0$}
\label{tab:dyn:2}
\begin{center}
\begin{tabular}{rccccccc}
  \hline\\
  & Mean ratio & Median ratio & S.D. & p .05 & p .10  & CS & K \\ 
  \hline\\ [3pt]
& \multicolumn{7}{c}{$N=100, T=8$}  \\ %[ 5 pt] 
   \cline{2-8} \\
   Infeasible & 1.317 &  &  &  &  &  &  \\
  ML & -0.307 & -0.323 & 0.024 & 0.827 & 0.884 & 36.7 &-  \\ 
  BC & 0.880 & 0.857 & 0.030 & 0.116 & 0.202 & 36.7 & -\\ 
  J & 0.657 & 0.655 & 0.032 & 0.347 & 0.452 & 36.7 & - \\ 
  GFE $\gamma$ = 0.1 & 0.752 & 0.753 & 0.028 & 0.094 & 0.156 & 18.8 & 32.408 \\ 
  GFE $\gamma$ = 0.4 & 1.072 & 1.069 & 0.029 & 0.066 & 0.112 & 10.7 & 12.872 \\ 
  GFE $\gamma$ = 0.7 & 1.178 & 1.168 & 0.029 & 0.069 & 0.118 & 8.2 & 8.352 \\ 
  GFE $\gamma$ = 1 & 1.248 & 1.233 & 0.029 & 0.086 & 0.127 & 6.6 & 6.262 \\   [ 8 pt]
  & \multicolumn{7}{c}{$N=200, T=8$}  \\ %[ 5 pt] 
   \cline{2-8} \\
    Infeasible & 1.315 &  &  &  &  &  &  \\ 
  ML & -0.339 & -0.346 & 0.018 & 0.981 & 0.985 & 36.6 & - \\ 
  BC & 0.850 & 0.846 & 0.022 & 0.150 & 0.237 & 36.6 & - \\ 
  J & 0.623 & 0.619 & 0.024 & 0.463 & 0.543 & 36.6 & - \\ 
  GFE $\gamma$ = 0.1 & 0.862 & 0.858 & 0.021 & 0.077 & 0.149 & 15.3 & 44.841 \\ 
  GFE $\gamma$ = 0.4 & 1.113 & 1.106 & 0.021 & 0.069 & 0.125 & 8 & 15.184 \\ 
  GFE $\gamma$ = 0.7 & 1.198 & 1.192 & 0.021 & 0.093 & 0.159 & 5.8 & 9.341 \\ 
  GFE $\gamma$ = 1 & 1.243 & 1.239 & 0.021 & 0.114 & 0.186 & 4.7 & 6.739 \\  [0.75cm]
& \multicolumn{7}{c}{$N=100, T=16$}  \\ %[ 5 pt] 
   \cline{2-8} \\
   Infeasible & 1.228 &  &  &  &  &  &  \\ 
 ML & 0.283 & 0.283 & 0.019 & 0.556 & 0.664 & 24.3 & - \\ 
  BC & 0.955 & 0.952 & 0.021 & 0.092 & 0.143 & 24.3 & - \\ 
  J & 0.895 & 0.888 & 0.022 & 0.223 & 0.311 & 24.3 & - \\ 
     GFE $\gamma$ = 0.1 & 0.814 & 0.820 & 0.020 & 0.092 & 0.151 & 13.7 & 37.497 \\ 
  GFE $\gamma$ = 0.4 & 1.006 & 1.014 & 0.021 & 0.060 & 0.121 & 8.3 & 15.770 \\ 
  GFE $\gamma$ = 0.7 & 1.071 & 1.078 & 0.020 & 0.063 & 0.115 & 6.5 & 10.285 \\ 
  GFE $\gamma$ = 1 & 1.120 & 1.125 & 0.021 & 0.070 & 0.122 & 5.5 & 7.762 \\  [8pt]
& \multicolumn{7}{c}{$N=200, T=16$}  \\ %[ 5 pt] 
   \cline{2-8} \\
   Infeasible & 1.224 &  &  &  &  &  &  \\ 
 ML & 0.261 & 0.258 & 0.013 & 0.847 & 0.908 & 24.1 & - \\ 
  BC & 0.932 & 0.931 & 0.014 & 0.085 & 0.156 & 24.1 & - \\ 
  J & 0.865 & 0.858 & 0.016 & 0.250 & 0.359 & 24.1 & - \\ 
  GFE $\gamma$ = 0.1 & 0.871 & 0.866 & 0.014 & 0.090 & 0.149 & 10.9 & 53.641 \\ 
  GFE $\gamma$ = 0.4 & 1.031 & 1.023 & 0.014 & 0.042 & 0.104 & 5.9 & 18.949 \\ 
  GFE $\gamma$ = 0.7 & 1.082 & 1.071 & 0.014 & 0.053 & 0.105 & 4.4 & 11.695 \\ 
  GFE $\gamma$ = 1 & 1.113 & 1.104 & 0.015 & 0.069 & 0.130 & 3.7 & 8.554 \\ 
    \hline 
\end{tabular}
\end{center}
\begin{footnotesize}
Notes: dynamic logit model. Mean and median of the ratio between estimated and
population APEs. S.D.: standard deviation of estimated APE. p.05 and p.10: empirical size of a two-sided $t$-test at the 0.05 and 0.1 significance level.  CS=\% of dropped observations, K= average number of groups in GFE estimators. 1000 Monte Carlo replications.
\end{footnotesize} 
\end{table}

\section{Empirical applications} \label{sec:applic}
\subsection{Female labor force participation}\label{sec:applic_labor}

We revisit the empirical application on inter-temporal labor supply
decisions of women, also illustrated in \cite{dj2015}. Data are related to the employment status of $N=1461$ married women aged between 18 and 60 in 1985, whose husbands were always employed in the period from 1981 to 1988, observed for $T=8$ years (PSID waves 15-22). We estimate a dynamic logit model and include control variables such as the number of kids of different ages, the logarithm of the yearly income of the husband, the age, and the age squared.

%__________________ cs ________________________________
The employment status exhibits strong inter-temporal correlation: 143 women are unemployed for the whole period, while 719 women are always employed. Therefore 862 units, that is around $60\%$ of the sample, do not exhibit any outcome variation and are dropped due to CS when a FE model is estimated. 
We should expect the GFE to keep increasingly more units as we increase the value of the hyper-parameter $\gamma$.

%____________________________

In Table \ref{app_lab} we compare the results obtained by the GFE approach with four alternative APE estimators: the plug-in pooled estimator, the FE plug-in ML estimator, the analytical bias corrected APE estimators by \cite{fernandez2009fixed} and the half-panel jackknife APE estimator by \cite{dj2015}. For what concerns GFE, we report estimates for $\gamma=0.4,0.6,1$ and the GFE APE estimates when $K$ is fixed and equal to 5.

The APEs for every variable across each estimation method are in line with the corresponding economic intuition, but the magnitude of the the APE for $y_{i,t-1}$ is rather different across estimators. 
The pooled estimator indicates a strong positive effect, which likely reflects an upward omitted variable bias from ignoring UH. In contrast, the FE ML estimator yields a much lower APE, which  may indicate a downward bias due to the IPP. Consequently, the  analytical and jackknife bias-corrected  estimators mitigate this issue, giving an estimated APE for $y_{i,t-1}$ twice the one obtained by the FE ML estimator.

In order to select the proper value of the GFE hyperparameter, we follow the proposed rule that implies $K <\sqrt{N}T \approx 305$ and $K\gg \sqrt{T}\approx 3$.
Out of the values of $\gamma$ giving rise to the estimates 
in Table \ref{app_lab}, only $\ga=0.6,1$ are compliant with the rule, while $\ga=0.4$ violates it. Also notice that, while greater than $\sqrt{T}$, $K=5$ is too close to the lower bound, as the results are identical to the pooled model, thus indicating that the number of approximating points is too small to guarantee a good description of the UH. When $\ga=0.4$, the approximation of the UH is likely to be sufficient but the estimated number of grouped FE is too large to control the IPP bias. In this vein, the choice of  $\gamma=0.6$, where the number of parameters is 1/4 with respect to FE estimation and the approximation of the UH is delivered by 233 support points, is suggested in this case.

\begin{figure}
    \centering 
    \includegraphics[width=0.85\textwidth]{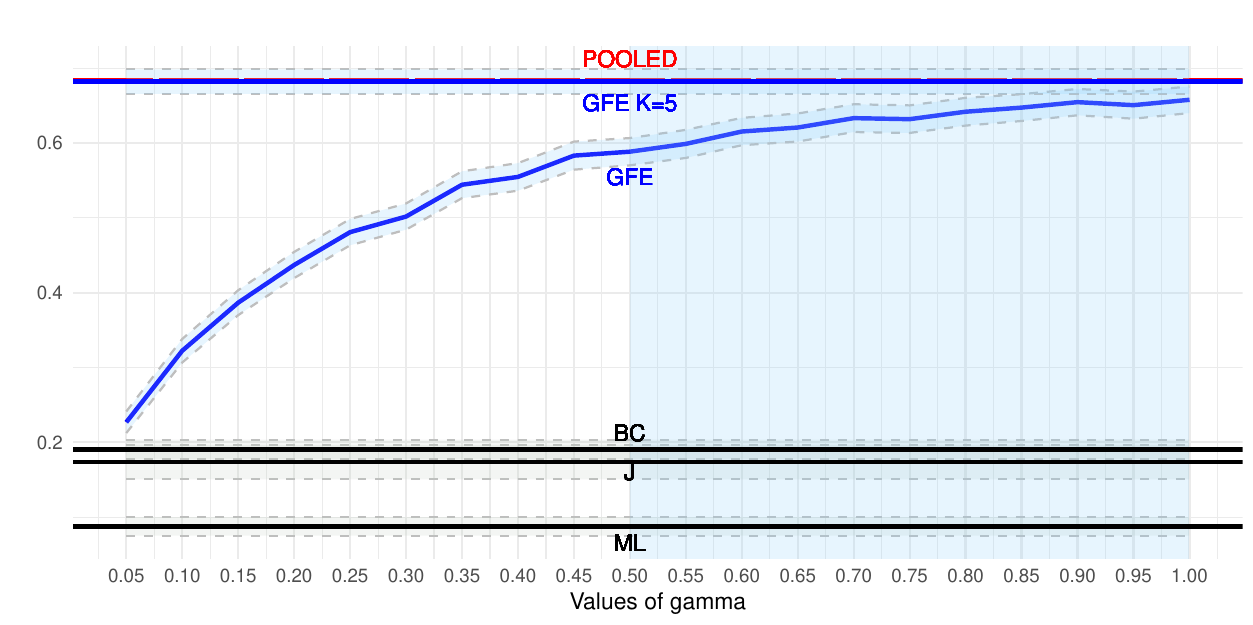} 
    \caption[Empirical application on labor market: graphical evaluation of the role of  $\ga$]{{\footnotesize Estimated value of the APE for $y_{i,t-1}$ plotted for 20 values of $\ga$. APE estimators: ``Pooled'' is the plug-in pooled estimator, ``ML'' is the FE plug-in ML estimator,``BC'' is the analytical bias corrected  estimators by \citep{fernandez2009fixed} and ''J'' is the half panel jackknife  estimator by \cite{dj2015}. "GFE" is the plug-in GFE APE estimator, "GFE K=5" is the plug-in GFE APE estimator with 5 groups. Lightblue area identifies values of $\ga$ which are compliant with the proposed rule.}}
    \label{plot_labor} 
\end{figure}

 As summarized by Figure \ref{plot_labor}, which reports
 the plug-in GFE APE estimator of $y_{i,t-1}$ for 20 values of $\gamma$, the GFE approach always gives a quantification of the effect that is greater than ML alternatives, in line with the findings in the  simulation study. Moreover, for increasing values of the hyperparameter, the plug-in GFE APE moves towards the plug-in pooled estimator, although in this case the number of groups with $\gamma=1$ is still sizable and equal to 117.

\begin{figure}
    \centering 
    \includegraphics[width=0.65\textwidth]{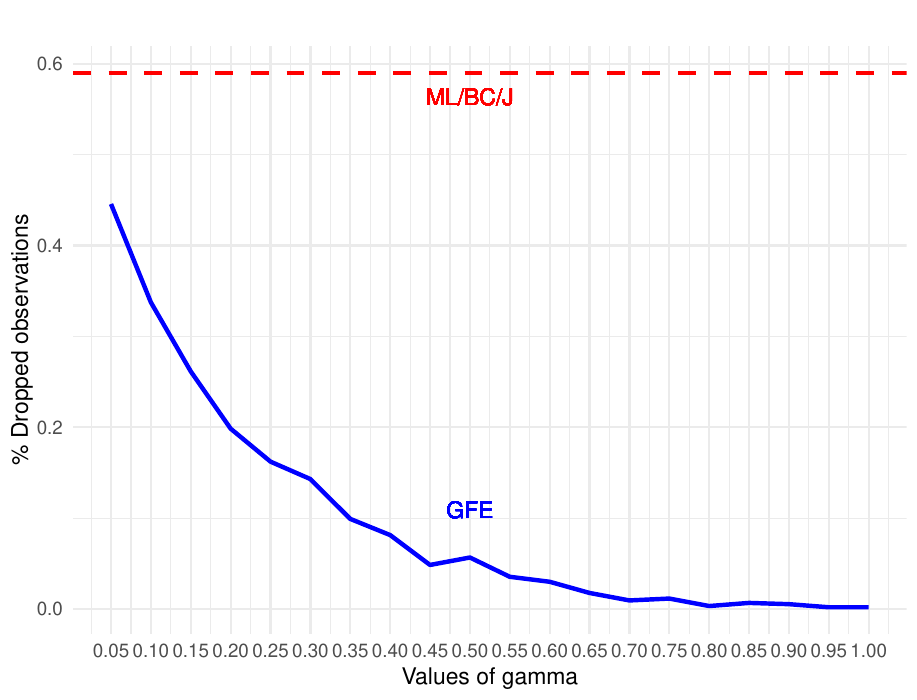} 
    \caption[Empirical application on labor market: percentage of dropped observations by estimator]{{\footnotesize Percentage of dataset dropped due to CS problem  plotted for 20 values of $\ga$. ``ML'' refers to FE estimator; ``BC'' is the bias corrected ML estimator by \cite{fernandez2009fixed},''J'' is the half panel jackknife \citep{dj2015}, ``GFE'' refers to the plug-in GFE estimator.}}
    \label{plot_labor_drop} 
\end{figure}

Regarding the GFE, the percentage of dropped
observations is  decreasing in $\gamma$ and the proposed approach stops dropping units for values of the hyperparameter higher than $0.8$. Figure \ref{plot_labor_drop} shows the decreasing trend of discarded units for increasing values of $\gamma$.

%__________________________TABELLA______________________
\begin{table}
\footnotesize
\begin{center}
\caption{ Empirical application on labor market: estimated APEs}
\label{app_lab}
\setlength{\tabcolsep}{3pt}
\begin{tabular}{lcccccccc}
\hline \\
 & Pooled & ML & BC & J & GFE $\ga=0.4$ & GFE $\ga=0.6$ & GFE $\ga=1$  & GFE $K=5$ \\[ 3 pt]
 \hline \\
$y_{t-1}$ & 0.684 & 0.088 & 0.191 & 0.173 & 0.555 & 0.616 & 0.658 & 0.682 \\ 
   & (0.008) & (0.006) & (0.007) & (0.012) & (0.009) & (0.009) & (0.009) & (0.008) \\[ 3 pt] 
  Child 0-2 & -0.051 & -0.069 & -0.076 & -0.098 & -0.039 & -0.040 & -0.035 & -0.051 \\ 
   & (0.007) & (0.008) & (0.008) & (0.014) & (0.009) & (0.009) & (0.009) & (0.007) \\[ 3 pt] 
  Child 3-5 & -0.010 & -0.032 & -0.031 & -0.049 & -0.002 & -0.005 & 0.002 & -0.012 \\ 
   & (0.007) & (0.007) & (0.007) & (0.015) & (0.009) & (0.009) & (0.009) & (0.007) \\[ 3 pt] 
  Child 6-17 & -0.001 & -0.012 & -0.012 & -0.021 & 0.001 & 0.002 & 0.005 & -0.002 \\ 
   & (0.003) & (0.006) & (0.006) & (0.012) & (0.006) & (0.006) & (0.006) & (0.003) \\ [ 3 pt]
  Inc. Husb & -0.019 & -0.026 & -0.032 & -0.037 & -0.030 & -0.023 & -0.025 & -0.020 \\ 
   & (0.004) & (0.007) & (0.007) & (0.011) & (0.008) & (0.008) & (0.007) & (0.004) \\[ 3 pt] 
  Age & 0.080 & 0.304 & 0.328 & 0.267 & 0.184 & 0.147 & 0.115 & 0.117 \\ 
   & (0.029) & (0.055) & (0.053) & (0.124) & (0.049) & (0.047) & (0.044) & (0.042) \\[ 3 pt] 
  $Age^2$ & -0.012 & -0.037 & -0.039 & -0.035 & -0.022 & -0.018 & -0.014 & -0.015 \\ 
   & (0.003) & (0.007) & (0.007) & (0.017) & (0.006) & (0.006) & (0.005) & (0.005) \\ [ 3 pt]
   \% Dropped & 0 & 59 & 59 & 59 & 8 & 3 & 0 & 0 \\ 
  K & 0 & 0 & 0 & 0 & 371 & 233 & 117 & 5 \\
  \hline
   \end{tabular}
\end{center}
\begin{footnotesize}
Standard errors (SE) in parentheses.  APE estimators: ``Pooled'' is the plug-in pooled estimator, ``ML'' is the FE plug-in ML estimator,``BC'' is the analytical bias corrected APE estimators by \citep{fernandez2009fixed} and ''J'' is the half-panel jackknife APE estimator by \cite{dj2015}. We report analytical SE for "Pooled", "ML" and "BC" estimators and Bootstrap SE for "J" based on $B=599$ replications. ``\% Dropped'' is the percentage  of
observations discarded for the CS problem. $K$ is the number of groups for individuals  found in the first step. $N=1461$, $T=8$.
\end{footnotesize}
\end{table}

 %%%%%%% FIGURA  1%%%%%%%

%%%%%%%      FIGURA 2    %%%%%%%%%%%
 
\subsection{An early warning system for banking crises}\label{sec:applic_crises}
%_____________ EWS ____________________
An early warning system for banking crises is a binary choice model where the
outcome variable takes value 1 if a banking crisis occurs in country
$i$ at time $t$ and 0 in non-critical periods. The probability of a
crisis is modeled as a function of lagged macroeconomic and financial
indicators that are supposed to warn about the likelihood of a crisis in advance. 
%_____________ DATASET _________________________
The dataset in exam, issued by \cite{laeven2018systemic}, consists of a balanced panel of $N=33$ countries
observed over the years 1986 - 2015 ($T=30$).
\cite{laeven2018systemic} give the definition of a banking crisis for a
large set of countries and identify 69 crisis episodes over 990 data
points, so we have a panel dataset where the dependent variable is an extremely rare event.
In addition to the one period lagged
dependent variable $y_{i,t-1}$, macroeconomic variables used in the
analysis, available as International Financial Statistics
(International Monetary Fund) or World Development Indicators
(World Bank) are: real GDP growth, the log of per capita GDP,
inflation, real interest rate, the ratio of M2 (broad money) to
foreign exchange reserves, the growth rate of real domestic credit and
the growth rate of foreign assets. All explanatory variables are
lagged by one period and further description of the dataset can be
found in \cite{pigini2021penalized} and \cite{caggiano2016comparing}.

\begin{figure}
    \centering 
    \includegraphics[width=1\textwidth, height=0.3\textheight]{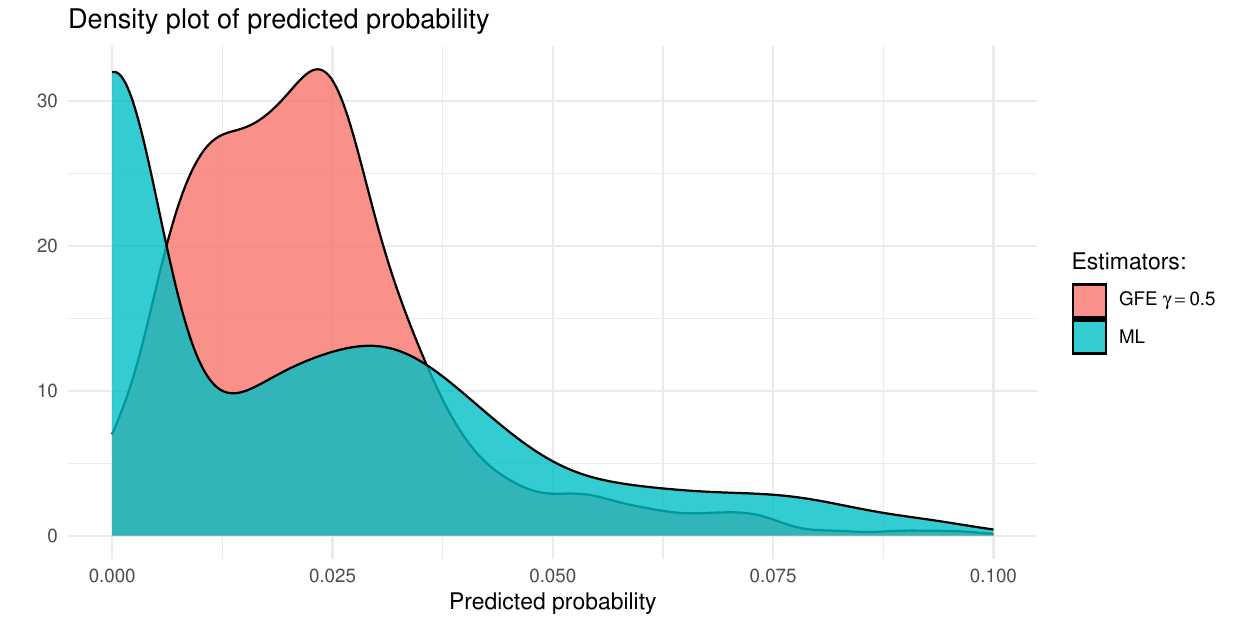} 
    \caption[Empirical application on banking crises: density estimation of predicted probability by estimator]{Empirical application on banking crises: density estimation of  in-sample predicted probability by estimator for the period 1986-2015.}
    \label{plot_density} 
\end{figure}

CS instances are a major issue in forecasting crises: in fact, the FE logit model cannot be used to predict the occurrence of a crisis for countries that never experienced one, as the estimates of the FEs  would not be finite.
In order to illustrate how the proposed approach can
circumvent this problem, Figure \ref{plot_density} depicts the
empirical density of in-sample predicted probability of crisis for
both ML and the GFE estimator ($\gamma=0.5$). The ML estimator drops 13 countries out of 33 due to CS and, as a result, we observe a large probability mass in 0. In contrast, the GFE approach drops only one country 
so that the empirical density of the forecast probability
turns out to be right-shifted compared
to the ML one, thus allowing non-trivial predictions of crises events for countries without outcome variation.

We also perform a one-step-ahead forecast exercise. Using an expanding training set stopping at years 2006 to 2010, we estimate a dynamic logit model and compute the out-of-sample predicted probabilities using the next year.
The last forecast year is 2011, as every year after that does not present any crisis in the dataset. The cut-offs used to compile confusion matrices are chosen by optimizing the in-sample sum of specificity and sensitivity. 
We compare the forecasting performance of the GFE approach to that of the FE ML estimator and analytical bias-corrected ML estimator \citep{fernandez2009fixed}. We experiment with 4 values of $\gamma=(0.005,0.1,0.5,1)$.

\begin{figure}
    \centering 
    \includegraphics[width=1\textwidth, height=0.3\textheight]{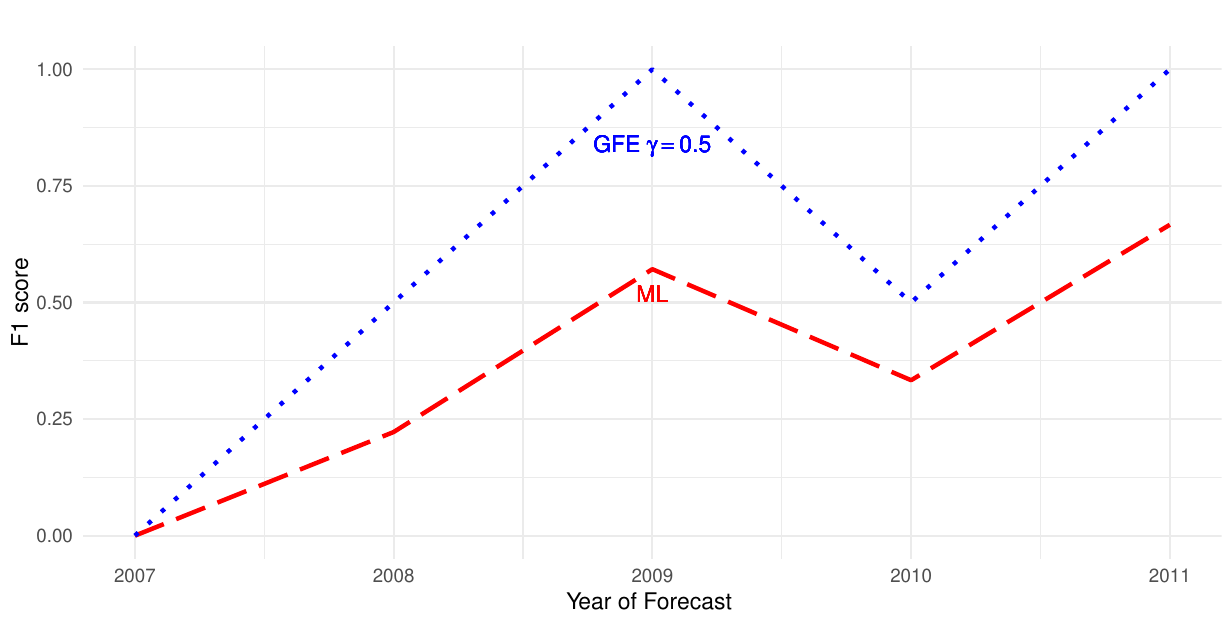} 
    \caption[Empirical application on banking crises: F1 score by estimator]{Empirical application on banking crises: in-sample F1 score by estimator for each forecast year.}
    \label{plot_f1} 
\end{figure}

Figure \ref{plot_f1} reports out-of-sample F1 score for ML and GFE
with $\gamma=0.5$ for all forecast years: the latter strictly
outperforms the former, achieving perfect classification in two out of
five scenarios (2009 and 2011). The better F1 score for GFE is
strictly due to the higher rate of false negatives detected. For the sake
of clarity, it is interesting to note that the number of groups
found by the GFE procedure with $\gamma=0.5$ in the first step varies in time
over the training sets - ranging from 7 to 9.

%_________ RESULTS____________________________-
The complete set of results of this forecast exercise is reported in Table \ref{app_crisis_forecast} in Appendix \ref{app:for}.
\begin{comment}
, where we report customary statistics such as the number of true positives, true negatives, false positives, false negatives and the F1 score together with the percentage of countries in CS and the number of groups for GFE.
Both  standard ML and GFE estimators manage to forecast a good amount of crises - 5 out of 7 in the 5 years  - although the GFE exhibits a better performance in  correctly predicting noncritical events.
Further considerations on these results
 should  take into account the cost of an undiagnosed crisis with respect to that of a false alarm, which is something we prefer not to take a stand on, since it is heavily dependent on policy-makers' priors.
\end{comment}
 Overall, the forecast performance of at least one of the GFE estimators considered in the exercise is always better than the one of the ML or BC estimators. In order to provide guidance on the choice of the hyperparameter, we suggest the choice of $\gamma=0.5$ as it is the value that gives the best performance in compliance with the proposed rule, since $\ga=1$, which would be slightly better in terms of F1 score, violates the lower bound.

\section{Conclusions}\label{sec:conclusions}

This paper motivates the use of the recently developed GFE approach  to perform regularized estimation of binary choice FE models in presence of severe CS. In such settings, FE models exhibit several deficiencies, including biased estimates of APE, inaccurate coverage, and the inability to generate meaningful predictions for units affected by CS.

We provide a simulation study concerning both static and dynamic specifications of logit models. Our results show that, by estimating a smaller number of FE, the proposed approach reduces the instances of CS and yields unbiased APE estimates with improved coverage properties, relative to the available alternatives. Moreover, by keeping all units grouped in clusters with response variability,  the GFE approach enables predictions for a much larger number of subjects in the sample.  

We also provide two illustrative examples, namely an analysis of determinants of labor force participation and a logit-based early warning system for rare bank crises. The first one shows that, by tuning the hyperparameter so as to provide a trade-off between a good approximation of the UH and a limited number of FEs to estimate, the GFE quantification of the APE for the lagged dependent variable diverges from the ML and bias corrected ones, while mitigating the potential omitted variable bias possibly exhibited by the pooled model. The second example focuses on forecasting and illustrates how the GFE approach manages to offer predictions for units that never experience a financial crisis in the training set.

\section*{Acknowledgments}
    We
    are grateful to Pavel \v{C}i\v{z}ek, Fulvio Corsi, Riccardo Lucchetti, Silvia Sarpietro, Laura Serlenga, Amrei
    Stammann, Rainer Winkelmann, to the participants of :
    the 12th Workshop of Econometrics and Empirical Economics, 13th IAAE annual conference, 30th IPDC, seminar at the  University of Pisa, for
    their helpful comments and suggestions.  Claudia Pigini and Alessandro Pionati would like to acknowledge the financial support by the European Union - Next Generation EU - Prin 2022, Project Code: 2022TZEXKF 02; Project CUP: I53D23002800006; Project Title: “Hidden Markov Models for Early Warning Systems”
    and financial coverage D.D. MUR 47/2025, CUP I33C25000280001; Project Title: ``The use of the Grouped Fixed Effects estimator in panel data analysis addressing unobserved heterogeneity”.
\begin{footnotesize}
\bibliography{biblio3.bib}

@article{hyslop1999state,
 title={State dependence, serial correlation and heterogeneity in intertemporal labor force participation of married women},
 author={Hyslop, Dean R},
 journal={Econometrica},
 volume={67},
 number={6},
 pages={1255--1294},
 year={1999},
 publisher={Wiley Online Library}
}

@article{fernandez2009fixed,
 title={Fixed effects estimation of structural parameters and marginal effects in panel probit models},
 author={Fern{\'a}ndez-Val, Iv{\'a}n},
 journal={Journal of Econometrics},
 volume={150},
 number={1},
 pages={71--85},
 year={2009},
 publisher={Elsevier}
}

@article{beyhum2024inference,
 title={Inference after discretizing unobserved heterogeneity},
 author={Beyhum, Jad and Mugnier, Martin},
 journal={arXiv preprint arXiv:2412.07352},
 year={2024}
}

@article{Kosmidis2009,
 title={Bias reduction in exponential family nonlinear models},
 author={Kosmidis, Ioannis and Firth, David},
 journal={Biometrika},
 volume={96},
 pages={793--804},
 year={2009},
 publisher={Oxford University Press}
}

@article{Heinze2002,
 title={A solution to the problem of separation in logistic regression},
 author={Heinze, Georg and Schemper, Michael},
 journal={Statistics in Medicine},
 volume={21},
 pages={2409--2419},
 year={2002},
 publisher={Wiley Online Library}
}

@article{Heinze2006,
 title={A comparative investigation of methods for logistic regression with separated or nearly separated data},
 author={Heinze, Georg},
 journal={Statistics in Medicine},
 volume={25},
 pages={4216--4226},
 year={2006},
 publisher={Wiley Online Library}
}

@article{Cook2018,
 title={Fixed effects in rare events data: {A} penalized maximum likelihood solution},
 author={Cook, Scott J and Hays, Jude C and Franzese, Robert J},
 journal={Political Science Research and Methods},
 volume = {2018},
 pages={1--14},
 year={2018},
 publisher={Cambridge University Press}
}

@article{kunz2021predicting,
  title={Predicting individual effects in fixed effects panel probit models},
  author={Kunz, Johannes S and Staub, Kevin E and Winkelmann, Rainer},
  journal={Journal of the Royal Statistical Society Series A: Statistics in Society},
  volume={184},
  number={3},
  pages={1109--1145},
  year={2021},
  publisher={Oxford University Press}
}

@article{hahnmoon2010,
 title={Panel data models with finite number of multiple equilibria},
 author={Hahn, Jinyong and Moon, Hyungsik Roger},
 journal={Econometric Theory},
 volume={26},
 number={3},
 pages={863--881},
 year={2010},
 publisher={Cambridge University Press},
}

@article{mugnier2025simple,
  title={A simple and computationally trivial estimator for grouped fixed effects models},
  author={Mugnier, Martin},
  journal={Journal of Econometrics},
  volume={250},
  pages={106011},
  year={2025},
  publisher={Elsevier}
}

@article{wang2024panel,
 title={Panel data models with time-varying latent group structures},
 author={Wang, Yiren and Phillips, Peter CB and Su, Liangjun},
 journal={Journal of Econometrics},
 volume={240},
 number={1},
 pages={105685},
 year={2024},
 publisher={Elsevier}
}

@article{lumsdaine2023estimation,
 title={Estimation of panel group structure models with structural breaks in group memberships and coefficients},
 author={Lumsdaine, Robin L and Okui, Ryo and Wang, Wendun},
 journal={Journal of Econometrics},
 volume={233},
 number={1},
 pages={45--65},
 year={2023},
 publisher={Elsevier}
}

@article{heckman1980does,
 title={Does unemployment cause future unemployment? Definitions, questions and answers from a continuous time model of heterogeneity and state dependence},
 author={Heckman, James J and Borjas, George J},
 journal={Economica},
 volume={47},
 number={187},
 pages={247--283},
 year={1980},
 publisher={JSTOR}
}

@article{contoyannis2004dynamics,
 title={The dynamics of health in the British Household Panel Survey},
 author={Contoyannis, Paul and Jones, Andrew M and Rice, Nigel},
 journal={Journal of Applied Econometrics},
 volume={19},
 number={4},
 pages={473--503},
 year={2004},
 publisher={Wiley Online Library}
}

@article{cappellari2004modelling,
 title={Modelling low income transitions},
 author={Cappellari, Lorenzo and Jenkins, Stephen P},
 journal={Journal of applied econometrics},
 volume={19},
 number={5},
 pages={593--610},
 year={2004},
 publisher={Wiley Online Library}
}

@article{alessie2004ownership,
 title={Ownership of stocks and mutual funds: A panel data analysis},
 author={Alessie, Rob and Hochguertel, Stefan and Soest, Arthur van},
 journal={Review of Economics and Statistics},
 volume={86},
 number={3},
 pages={783--796},
 year={2004},
 publisher={MIT Press 238 Main St., Suite 500, Cambridge, MA 02142-1046, USA journals~…}
}

@article{wooldridge2005simple,
 title={Simple solutions to the initial conditions problem in dynamic, nonlinear panel data models with unobserved heterogeneity},
 author={Wooldridge, Jeffrey M},
 journal={Journal of applied econometrics},
 volume={20},
 number={1},
 pages={39--54},
 year={2005},
 publisher={Wiley Online Library}
}

@article{pigini2016state,
 title={State dependence in access to credit},
 author={Pigini, Claudia and Presbitero, Andrea F and Zazzaro, Alberto},
 journal={Journal of Financial Stability},
 volume={27},
 pages={17--34},
 year={2016},
 publisher={Elsevier}
}

@article{bettin2018dynamic,
 title={A dynamic double hurdle model for remittances: evidence from Germany},
 author={Bettin, Giulia and Lucchetti, Riccardo and Pigini, Claudia},
 journal={Economic Modelling},
 volume={73},
 pages={365--377},
 year={2018},
 publisher={Elsevier}
}

@article{drescher2021determinants,
 title={Determinants, persistence, and dynamics of energy poverty: An empirical assessment using German household survey data},
 author={Drescher, Katharina and Janzen, Benedikt},
 journal={Energy Economics},
 volume={102},
 pages={105433},
 year={2021},
 publisher={Elsevier}
}

@article{arroyabe2022estimation,
 title={On the estimation of true state dependence in the persistence of innovation},
 author={Arroyabe, Marta F and Schumann, Martin},
 journal={Oxford Bulletin of Economics and Statistics},
 volume={84},
 number={4},
 pages={850--893},
 year={2022},
 publisher={Wiley Online Library}
}

@article{pigini2021penalized,
 title={Penalized maximum likelihood estimation of logit-based early warning systems},
 author={Pigini, Claudia},
 journal={International Journal of Forecasting},
 volume={37},
 number={3},
 pages={1156--1172},
 year={2021},
 publisher={Elsevier}
}

@article{caggiano2016comparing,
 title={Comparing logit-based early warning systems: Does the duration of systemic banking crises matter?},
 author={Caggiano, Giovanni and Calice, Pietro and Leonida, Leone and Kapetanios, George},
 journal={Journal of Empirical finance},
 volume={37},
 pages={104--116},
 year={2016},
 publisher={Elsevier}
}

@book{laeven2018systemic,
 title={Systemic banking crises revisited},
 author={Laeven, Mr Luc and Valencia, Mr Fabian},
 year={2018},
 publisher={International Monetary Fund}
}

@article{NS1948,
 title={Consistent estimates based on partially consistent observations},
 author={Neyman, Jerzy and Scott, Elizabeth L},
 journal={Econometrica},
 pages={1--32},
 year={1948},
 publisher={JSTOR}
}

@article{albert1984existence,
 title={On the existence of maximum likelihood estimates in logistic regression models},
 author={Albert, Adelin and Anderson, John A},
 journal={Biometrika},
 volume={71},
 number={1},
 pages={1--10},
 year={1984},
 publisher={Oxford University Press}
}

@article{freemanweidner2023,
 title={Linear panel regressions with two-way unobserved heterogeneity},
 author={Freeman, Hugo and Weidner, Martin},
 journal={Journal of Econometrics},
 volume={237},
 number={1},
 pages={105498},
 year={2023},
 publisher={Elsevier}
}

@article{dzemski2019,
 title={An empirical model of dyadic link formation in a network with unobserved heterogeneity},
 author={Dzemski, Andreas},
 journal={Review of Economics and Statistics},
 volume={101},
 number={5},
 pages={763--776},
 year={2019}
}

@article{hahn2011bias,
 title={Bias reduction for dynamic nonlinear panel models with fixed effects},
 author={Hahn, Jinyong and Kuersteiner, Guido},
 journal={Econometric Theory},
 volume={27},
 number={6},
 pages={1152--1191},
 year={2011},
 publisher={Cambridge University Press}
}

@article{Li2003,
 title={Efficiency of projected score methods in rectangular array asymptotics},
 author={Li, Haihong and Lindsay, Bruce G and Waterman, Richard P},
 journal={Journal of the Royal Statistical Society: Series B (Statistical Methodology)},
 volume={65},
 number={1},
 pages={191--208},
 year={2003},
 publisher={Wiley Online Library}
}

@article{bm2015,
 title={Grouped patterns of heterogeneity in panel data},
 author={Bonhomme, St{\'e}phane and Manresa, Elena},
 journal={Econometrica},
 volume={83},
 number={3},
 pages={1147--1184},
 year={2015},
 publisher={Wiley Online Library}
}

@article{blm2022,
 title={Discretizing unobserved heterogeneity},
 author={Bonhomme, St{\'e}phane and Lamadon, Thibaut and Manresa, Elena},
 journal={Econometrica},
 volume={90},
 number={2},
 pages={625--643},
 year={2022},
 publisher={Wiley Online Library},
}

@article{suppblm2022,
 title={Supplement to ``Discretizing unobserved heterogeneity''},
 author={Bonhomme, St{\'e}phane and Lamadon, Thibaut and Manresa, Elena},
 journal={Econometrica supplementary material},
 volume={90},
 number={2},
 pages={1--21},
 year={2022},
 publisher={Wiley Online Library},
}

@article{firth1993bias,
 title={Bias reduction of maximum likelihood estimates},
 author={Firth, David},
 journal={Biometrika},
 volume={80},
 number={1},
 pages={27--38},
 year={1993},
 publisher={Oxford University Press}
}

@article{HN2004,
 title={Jackknife and analytical bias reduction for nonlinear panel models},
 author={Hahn, Jinyong and Newey, Whitney},
 journal={Econometrica},
 volume={72},
 pages={1295--1319},
 year={2004},
 publisher={Wiley Online Library}
}

@article{phillips2016,
author = {Su, Liangjun and Shi, Zhentao and Phillips, Peter C. B.},
title = {Identifying Latent Structures in Panel Data},
journal = {Econometrica},
volume = {84},
number = {6},
pages = {2215-2264},
year = {2016},
}

@article{dj2015,
 title={Split-panel jackknife estimation of fixed-effect models},
 author={Dhaene, Geert and Jochmans, Koen},
 journal={The Review of Economic Studies},
 volume=82,
 number=3,
 pages={991--1030},
 year=2015,
 publisher={Oxford University Press}
}
\bibliographystyle{abbrvnat} % suggested by the Journal
\end{footnotesize}
\clearpage

\appendix 

\section{Additional simulation results}\label{app:sim}
    \begin{table}
   \caption{Estimated APE of $y_{t-1}$, $x_{it}=\alpha_i + 0.1(t - T/2) + N(0,1)$}
\label{tab:trending}
\vspace{-0.2cm}
\begin{center}
\begin{tabular}{rccccccc}
  \hline\\
  & Mean ratio & Median ratio & S.D. & p .05 & p .10  & CS & K \\ 
  \hline\\ [3pt]
& \multicolumn{7}{c}{$N=100, T=8$}  \\ %[ 5 pt] 
   \cline{2-8} \\
  Infeasible & 1.585 &  &  &  &  &  &  \\ 
ML & -0.378 & -0.396 & 0.022 & 0.770 & 0.843 & 49.6 & - \\ 
  BC & 0.852 & 0.829 & 0.027 & 0.149 & 0.217 & 49.6 & - \\ 
  J & 0.566 & 0.535 & 0.029 & 0.223 & 0.317 & 49.6 & - \\ 
  GFE $\gamma$ = 0.1 & 0.698 & 0.687 & 0.026 & 0.109 & 0.186 & 29.9 & 32.001 \\ 
  GFE $\gamma$ = 0.4 & 1.050 & 1.042 & 0.027 & 0.066 & 0.114 & 19.8 & 12.668 \\ 
  GFE $\gamma$ = 0.7 & 1.167 & 1.138 & 0.027 & 0.066 & 0.116 & 15.5 & 8.202 \\ 
  GFE $\gamma$ = 1 & 1.244 & 1.236 & 0.028 & 0.075 & 0.139 & 13 & 6.124 \\ [ 5 pt]
  & \multicolumn{7}{c}{$N=200, T=8$}  \\ %[ 5 pt] 
   \cline{2-8} \\
    Infeasible & 1.576 &  &  &  &  &  &  \\ 
ML & -0.399 & -0.394 & 0.016 & 0.956 & 0.976 & 49.2 &  \\ 
  BC & 0.836 & 0.840 & 0.020 & 0.175 & 0.240 & 49.2 &  \\ 
  J & 0.543 & 0.534 & 0.021 & 0.331 & 0.441 & 49.2 &  \\ 
  GFE $\gamma$ = 0.1 & 0.810 & 0.801 & 0.019 & 0.089 & 0.146 & 25.2 & 44.051 \\ 
  GFE $\gamma$ = 0.4 & 1.098 & 1.085 & 0.019 & 0.060 & 0.120 & 15.1 & 14.797 \\ 
  GFE $\gamma$ = 0.7 & 1.189 & 1.168 & 0.019 & 0.074 & 0.140 & 11.3 & 9.113 \\ 
  GFE $\gamma$ = 1 & 1.240 & 1.216 & 0.020 & 0.086 & 0.154 & 9.4 & 6.603 \\ [0.75cm]
& \multicolumn{7}{c}{$N=100, T=16$}  \\ %[ 5 pt] 
   \cline{2-8} \\
Infeasible & 1.431 &  &  &  &  &  &  \\ 
 ML & 0.189 & 0.185 & 0.016 & 0.567 & 0.669 & 36.7 & - \\ 
  BC & 0.901 & 0.899 & 0.018 & 0.095 & 0.152 & 36.7 & - \\ 
  J & 0.796 & 0.804 & 0.019 & 0.175 & 0.246 & 36.7 & - \\ 
  GFE $\gamma$ = 0.1 & 0.755 & 0.738 & 0.018 & 0.098 & 0.171 & 22.3 & 35.439 \\ 
  GFE $\gamma$ = 0.4 & 0.965 & 0.961 & 0.019 & 0.055 & 0.109 & 14.3 & 14.510 \\ 
  GFE $\gamma$ = 0.7 & 1.041 & 1.035 & 0.019 & 0.045 & 0.104 & 11.4 & 9.435 \\ 
  GFE $\gamma$ = 1 & 1.100 & 1.089 & 0.019 & 0.054 & 0.111 & 9.7 & 7.165 \\  [8pt] 

& \multicolumn{7}{c}{$N=200, T=16$}  \\ %[ 5 pt] 
   \cline{2-8} \\
 Infeasible & 1.429 &  &  &  &  &  &  \\ 
ML & 0.218 & 0.223 & 0.011 & 0.812 & 0.881 & 36.7 &  \\ 
  BC & 0.936 & 0.939 & 0.013 & 0.096 & 0.153 & 36.7 &  \\ 
  J & 0.831 & 0.816 & 0.014 & 0.184 & 0.280 & 36.7 &  \\ 
  GFE $\gamma$ = 0.1 & 0.861 & 0.857 & 0.013 & 0.091 & 0.154 & 19 & 50.007 \\ 
  GFE $\gamma$ = 0.4 & 1.029 & 1.030 & 0.013 & 0.052 & 0.102 & 11.1 & 17.279 \\ 
  GFE $\gamma$ = 0.7 & 1.088 & 1.080 & 0.013 & 0.057 & 0.105 & 8.4 & 10.629 \\ 
  GFE $\gamma$ = 1 & 1.135 & 1.128 & 0.013 & 0.067 & 0.122 & 0.071 & 7.798 \\ 
 \hline
\end{tabular}
\end{center}
\begin{footnotesize}
Notes: dynamic logit model with trending regressor. Mean and median of the ratio between estimated and
population APEs. S.D.: standard deviation of estimated APE. p.05 and p.10: empirical size of a two-sided $t$-test centered at the truth at the 0.05 and 0.1 significance level.  CS=\% of dropped observations, K= average number of groups in GFE estimators. 1000 Monte Carlo replications.
\end{footnotesize}
\end{table}

\newpage
\section{Empirical application on banking crises: full results}\label{app:for}

\begin{footnotesize}
\begin{longtable}{llllllll}
\caption{Empirical application on banking crisis: Forecast} \label{app_crisis_forecast}\tabularnewline
\hline\\
Forecast for 2007 & TRUE POS. & TRUE NEG.& FALSE POS. & FALSE NEG.& K &Drop &F1 \\ 
  \hline \\
  ML & 0 & 27 & 5 & 1 & - & 14 & - \\ 
  BC & 0 & 25 & 7 & 1 & - & 14 &  -\\ 
  GFE 0.005 & 0 & 25 & 7 & 1 & 31 & 13 & - \\ 
  GFE 0.1 & 0 & 32 & 0 & 1 & 18 & 10 & - \\ 
  GFE 0.5 & 0 & 32 & 0 & 1 & 7 & 8 & - \\ 
  GFE 1 & 0 & 32 & 0 & 1 & 5 & 1 & - \\ 
   \hline \\
Forecast for 2008 & TRUE POS. & TRUE NEG.& FALSE POS. & FALSE NEG.& K &Drop &F1  \\ 
  \hline \\
  ML & 1 & 25 & 6 & 1 & - & 14 & 0.222 \\ 
  BC & 1 & 25 & 6 & 1 & - & 14 & 0.222 \\ 
  GFE 0.005 & 1 & 30 & 1 & 1 & 30 & 13 & 0.5 \\ 
  GFE 0.1 & 1 & 30 & 1 & 1 & 19 & 9 & 0.5 \\ 
  GFE 0.5 & 1 & 30 & 1 & 1 & 8 & 1 & 0.5 \\ 
  GFE 1 & 1 & 31 & 0 & 1 & 5 & 1 & 0.667 \\ 
  \hline \\
   % \newpage %%%da muovere alla bisogna
Forecast for 2009 & TRUE POS. & TRUE NEG.& FALSE POS. & FALSE NEG.& K &Drop &F1  \\ 
  \hline \\
ML & 2 & 28 & 3 & 0 & - & 13 & 0.571 \\ 
  BC & 2 & 28 & 3 & 0 & - & 13 & 0.571 \\ 
  GFE 0.005 & 2 & 29 & 2 & 0 & 30 & 12 & 0.667 \\ 
  GFE 0.1 & 2 & 28 & 3 & 0 & 19 & 9 & 0.571 \\ 
  GFE 0.5 & 2 & 31 & 0 & 0 & 9 & 1 & 1 \\ 
  GFE 1 & 2 & 31 & 0 & 0 & 5 & 1 & 1 \\ 
   \hline \\ 
Forecast for 2010 & TRUE POS. & TRUE NEG.& FALSE POS. & FALSE NEG.& K &Drop &F1 \\ 
  \hline \\
  ML & 1 & 28 & 4 & 0 & - & 13 & 0.333 \\ 
  BC & 1 & 28 & 4 & 0 & - & 13 & 0.333 \\ 
  GFE 0.005 & 1 & 29 & 3 & 0 & 31 & 13 & 0.4 \\ 
  GFE 0.1 & 1 & 30 & 2 & 0 & 20 & 8 & 0.5 \\ 
  GFE 0.5 & 1 & 30 & 2 & 0 & 9 & 1 & 0.5 \\ 
  GFE 1 & 1 & 30 & 2 & 0 & 6 & 1 & 0.5 \\ 
   \hline\\ 
Forecast for 2011 & TRUE POS. & TRUE NEG.& FALSE POS. & FALSE NEG.& K &Drop&F1 \\ 
  \hline \\
  ML & 1 & 31 & 1 & 0 & - & 13 & 0.667 \\ 
  BC & 1 & 31 & 1 & 0 & - & 13 & 0.667 \\ 
  GFE 0.005 & 1 & 31 & 1 & 0 & 30 & 12 & 0.667 \\ 
  GFE 0.1 & 1 & 32 & 0 & 0 & 17 & 7 & 1 \\ 
  GFE 0.5 & 1 & 32 & 0 & 0 & 7 & 1 & 1 \\ 
  GFE 1 & 1 & 32 & 0 & 0 & 4 & 1 & 1 \\
   \hline
%   \end{tabular}
\end{longtable}
\end{footnotesize}
\begin{scriptsize}
``ML'' refers to ML estimator; ``BC'' is the bias corrected
  ML estimator by \cite{fernandez2009fixed}, ``GFE'' refers to the GFE estimator.
 ``TRUE POS.'' number of true positives, ``TRUE NEG.'' number of true negatives,
  ``FALSE POS.'' number of false positives, ``  FALSE NEG.'' number of false negatives.
  ``K'' is the number of groups for found in the first step, ``Drop'' is the number of countries dropped due to CS,
  ``F1'' is the out-of-sample F1 score.
  Optimal cut-off  for fitted probability chosen by maximizing the in-sample sum of specificity and sensitivity.
\end{scriptsize}

 \end{document}